\documentclass[preprint,12pt]{elsarticle}

\usepackage{amssymb}
\usepackage{amsmath}
\usepackage{multirow}    
\usepackage{makecell}
\usepackage{adjustbox}
\usepackage[version=4]{mhchem}
\usepackage{hyperref}
\usepackage{float}
\usepackage{xcolor}
\usepackage{comment}
\usepackage{soul}

\journal{Journal of Nuclear Materials}

\begin{document}

\begin{frontmatter}



\title{Evolution of Irradiation-Induced Defect Landscape through Edge Dislocation and Vacancy Loop Interactions in Tungsten} 

\author{Soumya Mishra and Suchandrima Das} 

\affiliation{organization={Department of Materials Engineering, Indian Institute of Science},   
            city={Bengaluru},
            postcode={560012}, 
            state={Karnataka},
            country={India}} 

\begin{abstract}

Structural materials in fusion reactors develop irradiation-induced defect populations that govern their mechanical behaviour. While advanced microscopy provides detailed characterisation of such defects, their dynamic evolution through dislocation-defect interactions during plastic deformation remains unresolved. Constitutive models may reproduce experimentally observed hardening even when the assumed representation of defect evolution is incomplete; therefore, such agreement does not necessarily establish the underlying atomistic mechanisms. Here, we address this gap using molecular dynamics simulations to resolve dislocation-defect interaction mechanisms and the resulting evolution of defect morphology and obstacle strength. Tungsten is the leading candidate plasma-facing material for fusion reactors, where vacancy loops are common irradiation defects. Using vacancy loops as a prototype defect, we systematically investigate their interactions with edge dislocations by varying loop size, orientation, and dislocation-loop intersection position. Parallel loops exhibit three distinct evolution pathways: complete annihilation, transformation into weaker remnant defects, or defect transport, depending on the intersection position and loop size. In contrast, inclined closed loops undergo dislocation reactions that produce sessile $\langle010\rangle$ dislocation segments, leading to substantially higher obstacle strengths with little sensitivity to the intersection position. Repeated dislocation interactions further demonstrate that the defect landscape is not static: remnant defects can persist, transform, or lose their ability to impede subsequent dislocations, depending on the preceding interaction mechanism. These results show that the obstacle strength of a vacancy loop is governed not only by its initial size and orientation, but also by its interaction history and the resulting defect structure. 

\end{abstract}

\begin{keyword}
Molecular dynamics \sep Vacancy loop \sep Dislocation–defect interactions \sep Irradiation hardening \sep Tungsten 
\end{keyword}

\end{frontmatter}



\section{Introduction}

Tungsten has emerged as the primary candidate for plasma-facing components in future fusion reactors owing to its high melting point, low sputtering yield, and excellent thermal conductivity \cite{knaster2016materials, abernethy2017predicting, wei2014first}. During service, these components are subjected to high-energy neutron bombardment and elevated temperatures, producing collision cascades that generate self-interstitial atoms and vacancies \cite{gilbert2015energy, rieth2013recent, das2019recent, beck2013effect, hasegawa2014neutron, zheng2010investigation}. Moreover, transmutation in pure tungsten results in the formation of gaseous elements such as hydrogen (H) and helium (He) (which diffuse in the material to various defects) \cite{hammond2017helium, debroglie2015temperature, lee2007hydrogen, lhuillier2011trapping} and also Rhenium (Re) and Osmium (Os) (resulting in Re- and Os-rich clusters) \cite{lloyd2024microstructural, gilbert2011neutron, gilbert2012integrated, tanno2007effects, you2017clustering, lloyd2019decoration}. These defects subsequently migrate, interact, and cluster to form a complex hierarchy of irradiation-induced defects spanning multiple length scales, from point-defect clusters to large dislocation loops and voids, whose morphology, density, and spatial distribution depend strongly on the irradiation dose, temperature, and irradiation source \cite{yi2016insitu, gilbert2008structure, hu2025new, hasegawa2014neutron, zenobia2012response, yi2018high, debelle2008first, ogorodnikova2013tem,bonny2020trends, chatzikos2022positron, ogorodnikova2019annealing, hu2025new}. 
 
This defect landscape has serious consequences on the mechanical properties of irradiated tungsten, giving rise to irradiation hardening, strain localisation, embrittlement, and the progressive degradation of structural integrity \cite{durrschnabel2021new, hardie2013effects, das2018effect, das2018helium, fang2018hydrogen, hofmann2015non, reza2020thermal, knaster2016materials, yi2013insitu,  rieth2013recent, debroglie2015temperature, makin1965model, makin1968obstacles, garrison2019mechanical, hu2016irradiation}. 
Nanoindentation experiments consistently demonstrate significant irradiation hardening arising from the increased resistance to dislocation motion \cite{das2018effect, das2019hardening, armstrong2013effects, das2019orientation, das2020modified, chen2023towards, shi2022experiments, lin2024atomic}. Likewise, studies using other techniques such as HR-EBSD (High-Resolution Electron Backscatter Diffraction) and HR-DIC (High-resolution Digital Image Correlation) reveal changes in lattice distortions, distribution of deformation fields, and evolution of geometrically necessary dislocations around nano-indents, as a function of irradiation (dose and source) \cite{birosca2019dislocation, das2020modified, guan2017crystal, edwards2022mapping}. Irradiation is further accompanied by the emergence of localised slip channels, indicating that deformation progressively modifies the surrounding defect population rather than occurring within a static defect landscape \cite{xu2021recent, das2019hardening, yadav2025characterisation, abernethy2019effects, song2024deformation}. 

Together, these experimental techniques provide a multiscale description of the irradiation-induced defect population and its consequences for mechanical behaviour. However, while experiments can characterise the defect population and measure the resulting mechanical response, they provide limited information on how the irradiation-induced defects evolves during plastic deformation and how individual defects interact with gliding dislocations. Mesoscale approaches such as crystal plasticity finite element (CPFE) modelling provide a framework for describing the deformation behaviour of irradiated materials and can reproduce experimentally observed mechanical responses through constitutive descriptions and calibration \cite{wang2004orientation, wang2020orientation, das2019orientation, shi2022experiments, lin2024atomic, erinosho2015strain, zan2023nanoindentation, xiao2019modelling, li2014predicting, roy2025crystal}. However, agreement between model predictions and experiments does not necessarily establish that the assumed underlying mechanisms are correct.

For example, a previous CPFE model for helium-implanted tungsten successfully reproduced the experimentally observed nanoindentation behaviour by assuming that irradiation hardening arose from the dissociation of helium from He-filled Frenkel pairs (\ce{HEe2V}-SIA complexes) followed by Frenkel pair recombination \cite{das2019hardening}. Subsequent molecular dynamics simulations, however, demonstrated that the governing interaction mechanism was fundamentally different: edge dislocations act as sinks for the self-interstitial atoms within these complexes and transport them away during unpinning, leaving behind helium-vacancy complexes \cite{das2024md}. Thus, although the earlier CPFE model reproduced the measured mechanical response, subsequent atomistic analysis revealed that the assumed evolution of the irradiation-induced defect structure was not the mechanism governing the dislocation-defect interaction. This example illustrates the value of atomistic simulations in establishing the physical basis of constitutive descriptions by resolving defect transformations and dislocation-defect reactions that cannot be identified from macroscopic mechanical response alone. Atomistic information on such mechanisms is therefore important for establishing physically meaningful descriptions of irradiated plasticity.

Atomistic modelling approaches have been widely used to investigate the underlying mechanisms—such as defect strength, defect absorption, dislocation reactions, the formation of remnant defect structures and local slip resistances \cite{rodney1999dislocation, shi2022experiments, xu2003molecular, ito2014molecular, guo2009pressure, das2024md, sandoval2015competing, saha2023microstructure, xie2026atomistic}. Previous molecular dynamics studies have provided important insights into dislocation interactions with several types of irradiation-induced defects in BCC metals. For example, in tungsten, MD simulations have shown that numerous nanoscale helium-vacancy complexes, although individually below the resolution limit of conventional microscopy, contribute to irradiation hardening through their cumulative resistance to dislocation motion \cite{ren2022revealing, das2024md}. Similarly, studies on the interaction between dislocations and helium bubbles in tungsten show that the interaction mechanism and obstacle strength vary with bubble size and He/V ratio \cite{ji2023atomistically}. Likewise, investigations of edge dislocation interactions with voids in tungsten have shown that voids impede dislocation motion through an Orowan-type bypass mechanism, with the pinning strength growing as the void size increases \cite{osetsky2021atomic, kazakov2024interaction}. However, studies by Osetsky et al. \cite{osetsky2021atomic} and Kazakov et al. \cite{kazakov2024interaction} reported different pinning strengths, arising from different strain rates and the lengths of the periodic dislocation segments used in their simulations. Thus, these studies highlight that the measured pinning strength depends not only on the defect but also on factors such as strain rate, and dislocation length, emphasising the importance of systematic comparisons performed under consistent simulation conditions.

Another important insight emerging from previous atomistic studies is that the relative orientation between a gliding dislocation and the irradiation-induced defect governs the interaction mechanism \cite{yu2024atomistic, bacon2006computer, liu2008molecular, terentyev2007effect, terentyev2008simulation, terentyev2010reactions, terentyev2013cr}. Simulations of $\frac{1}{2}\langle 111\rangle$ edge dislocations with $\frac{1}{2}\langle 111\rangle$-type interstitial loops in tungsten and iron have shown that these loops, with a Burgers vector parallel to the approaching dislocation, generally behave as relatively weak obstacles \cite{yu2024atomistic, bacon2006computer}. In contrast, when the Burgers vector of the loop is inclined to that of the dislocation, a dislocation reaction results in the formation of a sessile segment, which strongly pins the dislocation, as reported in both iron \cite{terentyev2007effect} and tungsten \cite{yu2024atomistic, lin2024atomic}. Similar orientation-dependent behaviour has also been reported for impurity-decorated loops, where hydrogen or rhenium segregation modifies the quantitative pinning strength without altering the governing interaction mechanism \cite{yang2023influence, ren2025unveiling}. Together, these studies establish that orientation and impurity segregation at the loop govern its interaction with a gliding dislocation. 

Recent efforts have begun to integrate atomistic dislocation–defect interaction mechanisms into crystal plasticity constitutive models, demonstrating the potential of atomistically informed multiscale frameworks for irradiated tungsten \cite{lin2024atomic}. However, developing a more complete understanding of irradiation hardening requires addressing a fundamental question: how does the irradiation-induced defect landscape evolve during plastic deformation? Previous atomistic studies have provided limited insight into how dislocation–defect interactions continuously modify both the defect and the dislocation, thereby changing the nature of subsequent interactions. This distinction is crucial because irradiation-induced defects are not static but dynamic obstacles whose morphology, stability, and interaction strength change with repeated dislocation interaction. Consequently, the mechanical behaviour of irradiated materials is governed not only by the strength of individual defects but also by the continuous evolution of the defect population itself. Capturing this evolution is therefore essential for developing physically meaningful constitutive descriptions of irradiation hardening and strain localisation.

Vacancy loops provide a suitable prototype system for investigating the evolution of irradiation-induced defects. Past studies have reported that, at low irradiation doses, elongated $\frac{1}{2}\langle111\rangle$ vacancy loops are the dominant defects in irradiated tungsten, which progressively grow and transform into larger polygonal configurations with increasing temperature \cite{yi2015characterisation, hasanzadeh2018three, mason2020observation, li2021revealing}. In particular, in-situ TEM studies have identified predominantly vacancy-type $\frac{1}{2}\langle111\rangle$ loops at very low irradiation doses, including 0.01 dpa \cite{mason2014elastic, yi2016insitu}. Vacancy clusters can exist as vacancy platelets (also referred to as open loops, although they are not true dislocation loops) and prismatic vacancy loops (closed loops) \cite{fikar2018nano, gilbert2008structure, yi2013insitu, yi2016insitu}. Their size- and morphology-dependent configurations make their mechanical stability and destruction pathways sensitive to dislocation shear, providing a controlled defect class in which the effects of defect size, orientation, intersection position, and repeated dislocation interaction can be systematically isolated. Vacancy loops are considered here as a prototype rather than a complete representation of the irradiation-induced defect landscape, which may vary with irradiation dose, temperature, and irradiation conditions. The stability and evolution of these vacancy-type defects are also temperature-dependent; therefore, the present study focuses on interactions at 300 K.

The interactions between vacancy loops and gliding dislocations in tungsten remain comparatively less explored. Understanding these interactions is important for establishing strengthening mechanisms and for determining how individual dislocation-defect interactions progressively modify the defect landscape during plastic deformation. Several questions remain unresolved. How do the underlying interaction mechanisms change with loop size? How does loop orientation influence the atomistic reactions governing dislocation pinning? To what extent does the position at which a dislocation intersects a loop determine the subsequent defect evolution? Most importantly, how do these interactions modify both the dislocation and the vacancy loop, thereby changing the obstacle encountered by subsequent dislocations and contributing to the continuous evolution of the defect landscape? Answering these questions is essential not only for understanding dislocation-loop interactions themselves but also for establishing how irradiation-induced defects evolve during plastic deformation.

In this work, we employ molecular dynamics simulations to elucidate the atomistic mechanisms governing the interaction of edge dislocations with vacancy loops in irradiated tungsten. The study focuses on edge dislocations because they provide an ideal model system for investigating the evolution of irradiation-induced defect structures. Although screw dislocations dominate plastic deformation in BCC metals owing to their high lattice friction and kink-pair controlled motion, their mobility is intrinsically limited by a comparatively high Peierls barrier \cite{ghafarollahi2021theory}. In contrast, edge dislocations exhibit much higher mobility and therefore experience a more pronounced reduction in velocity when interacting with irradiation-induced defects such as vacancy loops. Consequently, edge dislocations provide a particularly sensitive system for resolving the atomistic mechanisms governing defect evolution and obstacle strengthening. Furthermore, given the large number of simulations required to systematically investigate the effects of loop size, orientation, intersection position, and repeated dislocation interaction, the present work focuses on edge dislocations to establish the interaction mechanisms and evolution of obstacle strength of loops during repeated interactions with edge dislocations, which can form a basis for future studies involving mixed dislocations. The resulting interaction mechanisms and obstacle strengths should, therefore, be interpreted specifically in the context of the interaction of vacancy loops with edge dislocations; screw dislocations may exhibit distinct interaction mechanisms and strengthening behaviour and are not explicitly addressed in the present study.

Rather than treating vacancy loops as static obstacles, we investigate how individual dislocation-loop interactions modify both the loop and the dislocation, thereby driving the evolution of the irradiation-induced defect landscape during plastic deformation. By systematically examining the effects of loop morphology (vacancy platelets and prismatic loops), orientation, and the position of the dislocation-loop interaction, we identify the governing interaction mechanisms responsible for dislocation pinning and determine how these factors influence remnant defect formation following dislocation interaction. Furthermore, by investigating repeated interactions between unpinned dislocations and remnant defect configurations, we demonstrate how successive dislocation-defect interactions progressively modify the defect landscape, thereby altering subsequent dislocation interactions. These results provide atomistic insight into the evolution of obstacles (vacancy loops) during plastic deformation and highlight the importance of accounting for defect transformation and interaction history when interpreting irradiation-induced hardening.

\section{Simulation method}
\label{section_simulation_method}

\begin{figure}[H]
\centering
\includegraphics[width = 1.0\linewidth]{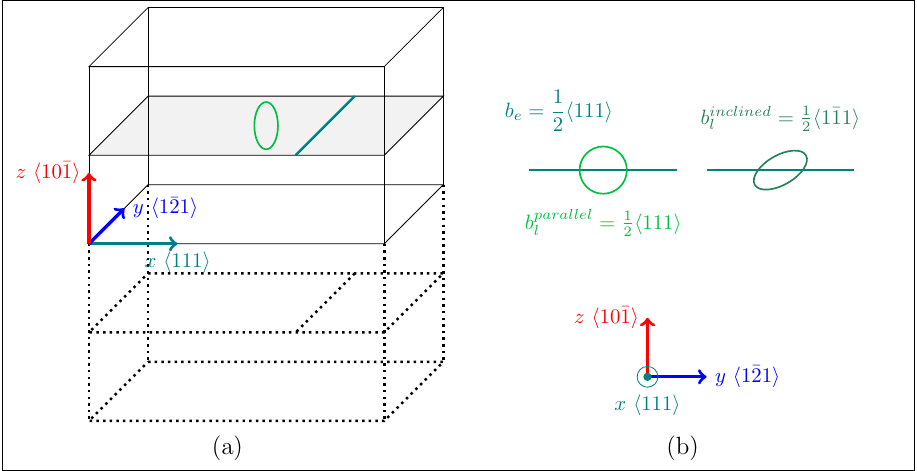}
\caption{(a) Schematic of the simulation box containing an edge dislocation and the vacancy loop; the grey plane is the glide plane of the edge dislocation ($b_e = 1/2\langle111\rangle$). (b) Orientation of the vacancy loop (with Burgers vector $b_l$) with respect to the edge dislocation ($b_e$). The X-axis is out of the plane of the paper.}
\label{figure_schematic_simulation_box_loop_dislocation}
\end{figure}

Details of the simulation setup used here can be found in  \cite{das2024md}. Simulations in this study were performed using the interatomic potential developed by Bonny et al. \cite{bonny2013mobility}. Briefly, the simulations are set up as follows: The simulation box was built with the lattice parameter $a0 = 0.315$ nm. The lattice parameter was obtained by relaxing a pure tungsten cell at 300 K with the chosen potential. In a simulation box of size $50 \times 40 \times 60  \thickspace nm^3$ (oriented along $\langle1 1 1\rangle$, $\langle1 \bar2 1\rangle$), and $\langle1 0 \bar1\rangle$ respectively along X, Y and Z directions), we create an edge dislocation dipole (with line direction along Y and Burgers vector along X) 10 nm from the centre along X direction (Figure~\ref{figure_schematic_simulation_box_loop_dislocation} (a)).  Then we relax the configuration under periodic boundary conditions and cut the simulation box in half along the Z-direction. Thus, we retain a negative edge dislocation with an extra half-plane below the slip plane within a box containing approximately 3.6 million atoms. The resulting dislocation density is about $5 \times 10^{14} \thickspace m^{-2}$. This is of the same order as the average GND density measured around a nano-indent of self-ion-implanted tungsten at a damage level of 0.01 dpa \cite{das2020modified}, providing a physically relevant reference for the magnitude of the dislocation-density scale considered in the present simulations. Simulations corresponding to this dislocation density require about 2000 to 3000 CPU hours per pass of the dislocation through the cell. To model a higher GND density of the order of $\times 10^{17} \thickspace m^{-2}$, the simulation cell would have to be so small that image forces would become prominent and loops of all sizes could no longer be modelled reliably. Conversely, modelling a lower dislocation density would require a much larger simulation box, leading to prohibitive computational costs. The simulation-cell dimensions were therefore guided by the physically relevant dislocation-density scale described above and chosen as a practical compromise between minimising the influence of boundary and image-force effects, maintaining a sufficiently large computational domain to represent vacancy loops over the considered size range, and ensuring computational feasibility.

After introducing the dislocation, we create vacancy loops of different sizes and orientations at the centre of the simulation box (Figure~\ref{figure_schematic_simulation_box_loop_dislocation}(a)) using Atomsk \cite{hirel2015atomsk}. We consider loop sizes of 1 to 10 nm reported in low-dose (0.01 dpa) irradiation experiments at room temperature by Yi et al. in tungsten \cite{yi2015characterisation}. 

Figure~\ref{figure_schematic_simulation_box_loop_dislocation} (a) shows a schematic of the simulation cell. The grey plane represents the glide plane of the $1/2\langle111\rangle$ edge dislocation and $1/2\langle111\rangle$ vacancy loop positioned in its path. Figure~\ref{figure_schematic_simulation_box_loop_dislocation} (b) shows the two orientations of the vacancy loop with different Burgers vectors relative to the $1/2\langle111\rangle$ edge dislocation that will be considered in subsequent simulations. Table~\ref{table_burgers_vector_loop_dislocation} lists the Burger vectors of vacancy loops ($b_l$) and edge dislocation ($b_e$). We note that the schematic presents only the case where the edge dislocation meets the centre of the vacancy loop; however, the top and bottom interactions are also presented in this paper. The interaction mechanisms in the simulations were visualised with OVITO \cite{stukowski2010ovito}.  

A periodic boundary condition is applied along the X and Y directions. Along Z direction, a few top layers of the simulation box are sheared (keeping a few layers at the bottom fixed) with a constant strain rate $10^{-5} \text{ ps}^{-1}$ ($10^{7} \text{ s}^{-1}$) at 300 K, thereby driving the edge dislocation towards the loop to interact with it. 

The force response associated with the dislocation motion was obtained from the interaction force between the sheared top boundary and the mobile atoms using the group/group calculation in LAMMPS. The $x$-component of this force, $F_{\mathrm{int},x}$, corresponding to the direction of dislocation motion, was extracted at each timestep. The force obtained from the simulation is expressed in eV/\AA\ and was converted to $\mu$N using the appropriate energy and length conversion. The resulting force-time curves are used to quantify the resistance the dislocation encounters as it passes through the simulation cell. The corresponding defect-free simulation, referred to as the “no-loop” case, provides the intrinsic lattice resistance under the same loading conditions. For the loop-containing simulations, we define the pinning strength as the maximum force peak observed in the force-time curve. This peak-based measure is used consistently to compare the obstacle strength of different vacancy loops.

\begin{table}[H]
    \centering
    \begin{tabular}{|c|c|c|c|}
    \hline
    Loop orientation & $b_l$ & $b_e$ & angle\\
    \hline
        parallel & $\frac{1}{2}\langle 111\rangle$     & \multirow{2}{*}{$\frac{1}{2}\langle 111\rangle$} & $0^{\circ}$ \\
        inclined & $\frac{1}{2}\langle 1\bar11\rangle$ & &  $70.53^{\circ}$\\
    \hline    
    \end{tabular}
    \caption{Burger vectors of vacancy loops ($b_l$) and edge dislocation ($b_e$) and the angle between them}
    \label{table_burgers_vector_loop_dislocation}
\end{table}

A strain-rate sensitivity analysis was performed to assess the influence of the imposed loading rate on the extracted interaction strength. The selected strain rate of $10^{-5}\,\mathrm{ps^{-1}}$ provides a practical means of driving the dislocation through the obstacle and directly extracting the maximum interaction force from the force response, without requiring a separate stress-threshold search for each loop size, orientation, and intersection configuration. Lowering the strain rate below this value produced only marginal changes in the extracted pinning force, as shown in Figure~\ref{figure_dislocation_F_vs_strain_rate} in Appendix~\ref{appendix_strain_rate_dependence}, while incurring significantly higher computational costs. We therefore use $10^{-5}\,\mathrm{ps^{-1}}$ for the systematic simulations presented here.

\begin{table}[H]
    \centering
    \begin{tabular}{|c|c|c|}
    \hline
    Interaction & Loop & Dislocation\\
    \hline
        A & Pristine & Pristine \\
        B & Pristine & Unpinned \\
        C & Remnant  & Pristine \\
        D & Remnant  & Unpinned \\
    \hline    
    \end{tabular}
    \caption{Types of interaction possible between a dislocation loop and an edge dislocation. We consider only interaction types A and D in this paper. Interaction A involves a pristine dislocation and a pristine loop, whereas Interaction D involves an unpinned dislocation and a remnant loop, representing an interaction between two objects that have each undergone a previous dislocation-loop interaction. In the present simulations, Interaction D is obtained by allowing the unpinned dislocation to move through the simulation box a second time along the periodic X-direction.}
    \label{table_loop_dislocation_interaction}
\end{table}

Table~\ref{table_loop_dislocation_interaction} summarises the possible interactions between a dislocation loop and an edge dislocation. Here, a pristine loop'' refers to an irradiation-induced loop that has not previously interacted with a dislocation, whereas a remnant loop'' refers to the loop structure remaining after a dislocation has interacted with the pristine loop. Similarly, a pristine dislocation'' is a dislocation that has not previously interacted with a loop, whereas an unpinned dislocation'' refers to the dislocation after it has interacted with a pristine loop and may consequently have undergone structural modification.

Out of the four possible interactions listed in Table~\ref{table_loop_dislocation_interaction}, we focus exclusively on interaction types A and D to investigate the evolution of dislocation-loop interactions. Interaction A represents the initial interaction between a pristine dislocation and a pristine loop. Following this interaction, the dislocation becomes unpinned and may undergo structural modification, while the loop is transformed into a remnant loop. Thus, an A-type interaction produces two modified objects: an unpinned dislocation and a remnant loop. In an irradiated defect landscape containing multiple dislocations and vacancy loops, a subsequent interaction need not involve the same dislocation and the same loop. For example, one dislocation may first interact with one pristine loop and become an unpinned, modified dislocation, while another dislocation may interact with another pristine loop and leave behind a remnant loop. The modified dislocation from the first interaction may then encounter the remnant loop produced by the second interaction. Interaction D corresponds to this situation, namely, the interaction between a dislocation that has previously experienced a dislocation–loop interaction and a loop that has likewise already undergone a prior dislocation–loop interaction. The two interacting objects, therefore, do not necessarily originate from the same preceding interaction.

In the present simulations, this interaction is achieved by allowing the unpinned dislocation produced by Interaction A to pass through the simulation box a second time due to the periodic boundary in the X-direction, thereby interacting with the remnant loop. The second pass is therefore used to obtain an interaction between the unpinned dislocation and the remnant loop in a controlled atomistic simulation; it is not intended to imply that the same dislocation repeatedly encounters the same physical remnant loop in an irradiated material.

\subsection{Vacancy loop configurations}
The vacancy loops considered in the present simulations relax into two distinct configurations depending on their size and orientation: vacancy platelets and closed prismatic vacancy loops. Following the terminology commonly used in the literature, the vacancy platelets are also referred to as ``open loops'' \cite{gilbert2008structure}, although they are not true (closed) dislocation loops. Gilbert et al. \cite{gilbert2008structure} describe these open loops as planar voids, formed by the removal of atoms from successive atomic planes; their formation energy is dominated by the surface energy and scales linearly with the number of vacancies ($N$). In contrast, the ``closed loops'' are collapsed dislocation loops whose energy includes contributions from the elastic field and the dislocation core. The relative stability of these two configurations is strongly size dependent. For $1/2\langle111\rangle$ vacancy defects in tungsten, Gilbert et al. \cite{gilbert2008structure} report that the open configuration is energetically favoured at smaller sizes, whereas the closed configuration becomes more stable above a critical diameter of approximately 3.4 nm. In the present simulations, the vacancy defects likewise form platelet (open) or closed loop configurations depending on defect size and orientation, as discussed in the following sections. Representative atomistic configurations of the vacancy platelet (open loop) and closed loop configurations considered in this work are shown in Figure~\ref{figure_parallel_centre_d_1nm_top_centre_first_pass_pinning_force} (a) and Figure~\ref{figure_parallel_5nm_centre_first_second_pass_pinning_force} (a), respectively. 

\section{Results and Discussion}

In order to investigate the effect of the relative orientation between the edge dislocation and the vacancy loop on the interaction mechanism, two types of orientations will be discussed here. We first consider parallel vacancy loops, in which the Burgers vectors of the dislocation and the loop are parallel. For parallel loops, variations in loop morphology, intersection position, and repeated dislocation interaction are examined to assess their influence on the evolution of the defect landscape. Furthermore, for all subsequent results, the force-time curves presented below correspond to the $x$-component of the inter-group force between the sheared top boundary and the mobile atoms, obtained from the MD simulations. Since this is a signed force component, its positive and negative values indicate opposite directions of the force response along the dislocation glide direction; the sign itself is not used to define a separate hardening contribution. For all subsequent results, we refer to the maximum force observed in these force-time curves as the pinning strength. The defect-free (without loop) curve provides the reference intrinsic lattice resistance under the same conditions (see, for example, green curve in Figure~\ref{figure_parallel_centre_d_1nm_top_centre_first_pass_pinning_force} (d)). The additional resistance observed in the presence of loops arises due to dislocation–loop interaction (see, for example, red curve in Figure~\ref{figure_parallel_centre_d_1nm_top_centre_first_pass_pinning_force} (d)).

\subsection{Parallel loop}
\label{section_parallel}

\subsubsection{Dislocation interaction with vacancy platelets (open loops)}

\begin{flushleft}
\textit{Dislocation intersects centre of platelet}
\end{flushleft}

\begin{flushleft}
\textbf{\small Interaction A}
\end{flushleft}

\begin{figure}[H]
\centering
\includegraphics[width = 0.8\linewidth]{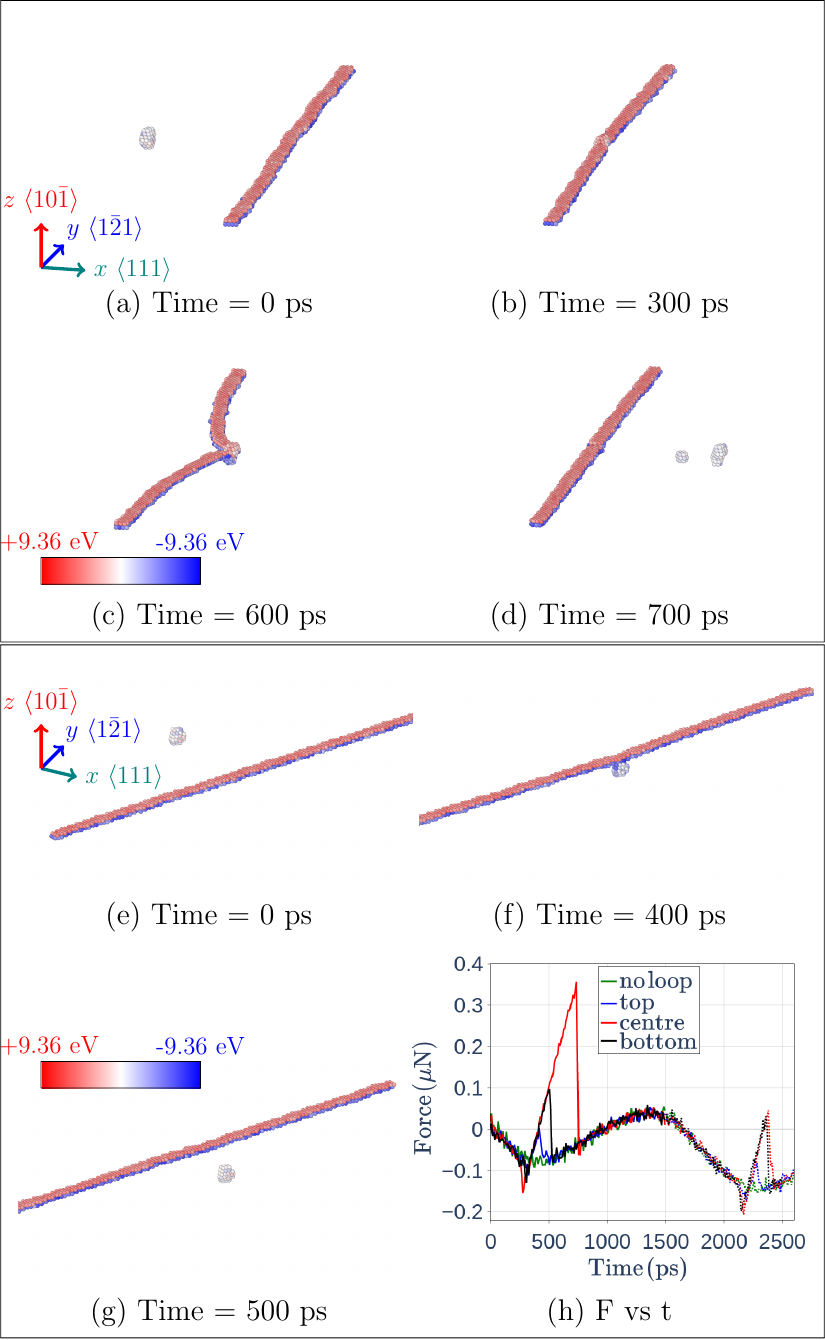}
\caption{Different stages of the interaction of an $\frac{1}{2}\langle 111\rangle$ edge dislocation with a $\frac{1}{2}\langle 111\rangle$ vacancy loop of size 1 nm, when the dislocation meets the loop at the centre,  during the first (interaction type A, (a)-(d)) and second pass (interaction type D, (e)-(h)) of the dislocation through the simulation cell. The corresponding force \textit{vs.} time curve is shown as the solid red curve in (h). The colouring of the atoms indicates the stress per atom in the x-direction (in eV).}
\label{figure_parallel_centre_d_1nm_top_centre_first_pass_pinning_force}
\end{figure}

Figure~\ref{figure_parallel_centre_d_1nm_top_centre_first_pass_pinning_force} (a-d) shows the different stages of the interaction of a $\frac{1}{2}\langle 111\rangle$ edge dislocation meeting the centre of a $\frac{1}{2}\langle 111\rangle$ vacancy loop of size 1 nm. At this size, a vacancy platelet (or open loop) is more stable than a closed loop, which is in agreement with Gilbert et al. \cite{gilbert2014comparative}, who report that below 3.4 nm, open loops are energetically favourable. The edge dislocation moves towards the vacancy platelet and is pinned by it. As stress increases over time, the dislocation bows and eventually breaks away. In this paper, such interaction will be referred to as ``Interaction I'' (see Table~\ref{table_loop_dislocation_interaction_mechanism}. The corresponding force-time curve is shown in solid red in  Figure~\ref{figure_parallel_centre_d_1nm_top_centre_first_pass_pinning_force} (h). The pristine loop is slightly altered in configuration, and only some vacancies are strewn in the wake of the unpinned dislocation.

\begin{flushleft}
\textbf{\small Interaction D}
\end{flushleft}

The interaction of the unpinned dislocation with the remnant vacancy loop is shown in Figure~\ref{figure_appendix_parallel_centre_d_1nm_second_pass} in  Appendix~\ref{appendix_parallel_interaction_mechanism}. The unpinned dislocation exits from one side of the simulation cell and enters from the other side, owing to the periodic boundary condition in the X-direction. The associated pinning force \textit{vs.} time curve (dotted red curve in Figure~\ref{figure_parallel_centre_d_1nm_top_centre_first_pass_pinning_force} (h)) shows a second, much smaller peak. This may be attributed to the smaller size of the remnant defect (left behind after the first run through the simulation box), which has fewer vacancies than the pristine version of the defect, as some were dislodged by the breaking-away dislocation during unpinning (see Figure~\ref{figure_parallel_centre_d_1nm_top_centre_first_pass_pinning_force} (d)). 

However, the rest of the curve almost perfectly overlaps with the green curve, which represents the inherent lattice resistance experienced by the dislocation as it moves through a cell in the absence of a loop. Thus, a dislocation intersecting at the centre deforms the platelet, generating a remnant obstacle that offers negligible resistance to subsequent dislocations. 

\begin{table}[H]
    \centering
    \begin{tabular}{|c|c|}
    \hline
    Interaction & Description \\
    Mechanism   &   \\
    \hline
        Type I & Dislocation is pinned by the loop, it bows and breaks away \\
     \hline    
        Type II & $\langle010\rangle$ segment forms, which pins the gliding dislocation, \\
          & the free segments bow out and break away \\
    \hline    
    \end{tabular}
    \caption{Types of frequently observed interaction mechanisms between a dislocation loop and an edge dislocation in this work. This list is intended only for ease of reference and is not an exhaustive classification of all dislocation–vacancy loop interaction mechanisms observed in this study.}
    \label{table_loop_dislocation_interaction_mechanism}
\end{table}

\begin{flushleft}
\textit{Dislocation intersects top of platelet}
\end{flushleft}

Figure~\ref{figure_parallel_centre_d_1nm_top_centre_first_pass_pinning_force}(e-g) shows the different stages of the interaction of a $\frac{1}{2}\langle 111\rangle$ edge dislocation meeting the top of a $\frac{1}{2}\langle 111\rangle$ vacancy loop of size 1 nm. Similar interaction mechanisms were observed when the dislocation meets the top of the loop. In contrast to the centre interaction, the top intersection results in only a negligible, nearly imperceptible pinning (blue curve in Figure~\ref{figure_parallel_centre_d_1nm_top_centre_first_pass_pinning_force}(h)). The reduced cross-section area of interaction between the gliding dislocation and the vacancy platelet limits the extent of dislocation–defect interaction, allowing the dislocation to move with negligible resistance from the defect (Figure~\ref{figure_parallel_centre_d_1nm_top_centre_first_pass_pinning_force}(f)). Consequently, no measurable increase in obstacle strength is observed during either the initial (Interaction A) or subsequent dislocation interaction (Interaction D), as can be seen in Figure~\ref{figure_parallel_centre_d_1nm_top_centre_first_pass_pinning_force} (h) (the blue curve almost perfectly overlaps with the green curve, which represents the inherent lattice resistance experienced by the dislocation as it moves through a cell without any loop present).

\begin{flushleft}
\textit{Dislocation intersects bottom of platelet}
\end{flushleft}

When the dislocation meets the bottom of the loop, the interaction mechanism (``Interaction I'') is the same as that observed during centre intersection, during both initial (Interaction A) and subsequent dislocation interaction (Interaction D). However, bottom intersection results in a much lower pinning force in the first pass (Interaction A), similar to the top interaction, as can be seen in the black curve in Figure~\ref{figure_parallel_centre_d_1nm_top_centre_first_pass_pinning_force}(h). This may be due to the reduced cross-sectional area of interaction between the gliding dislocation and the vacancy platelet, similar to the top interaction. During the second pass (Interaction D), the pinning strength of the bottom intersection is the same as that of the centre intersection.

In summary, these results show that the interaction mechanism of the 1 nm vacancy platelet is governed primarily by the intersection position of the loop with the dislocation. Only dislocations intersecting the centre of the loop result in significant pinning, which is substantially eliminated during the second pass due to structural modification of the platelet following the initial interaction. In contrast, the off-centre interactions result in significantly reduced (bottom) or no pinning (top) and therefore contribute little to the obstacle strength. These observations indicate that small vacancy platelets act as weak, highly geometry-sensitive obstacles whose pinning strength weakens following repeated dislocation interactions. The behaviour changes for larger loops, where the transition to closed loops leads to entirely different interaction mechanisms depending on the intersection position, as described in the following section. 

\subsubsection{Dislocation interaction with closed loops}

\begin{flushleft}
\textit{Dislocation intersects centre of loop}
\end{flushleft}

\begin{flushleft}
\textbf{\small Interaction A}
\end{flushleft}

\begin{figure}[H]
\centering
\includegraphics[width = 0.8\linewidth]{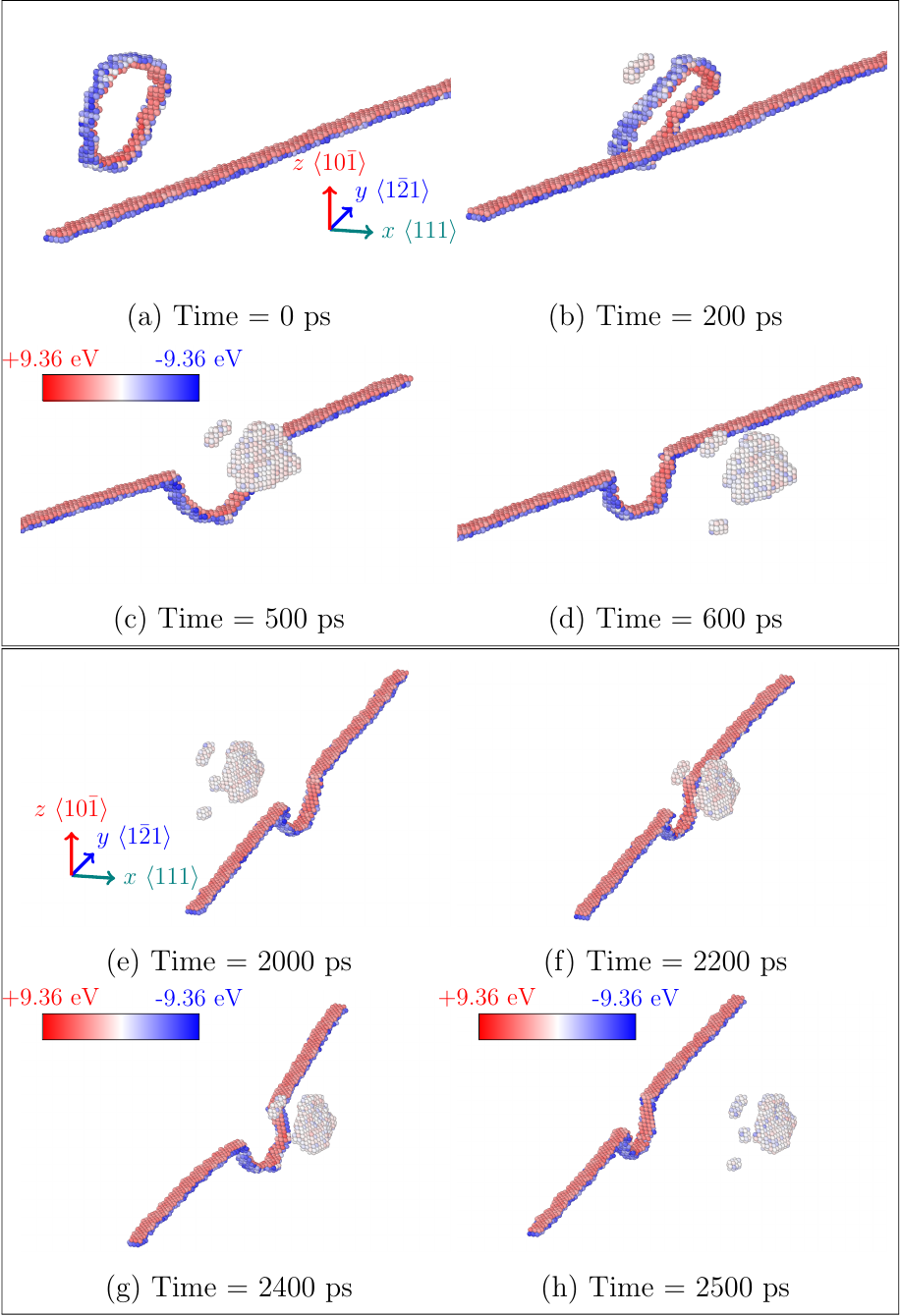}
\caption{Different stages of the interaction of an $\frac{1}{2}\langle 111\rangle$ edge dislocation with a $\frac{1}{2}\langle 111\rangle$ vacancy loop of size 5 nm, when the dislocation meets the loop at the centre, during the first (interaction type A, (a)-(d)) and second pass (interaction type D, (e)-(h)) of the dislocation through the simulation cell. The corresponding curve of force (averaged over 10 ps) \textit{vs.} time is shown in red in Figure~\ref{figure_parallel_top_bottom_d_5nm_first_pass_pinning_force}(d). The colouring of the atoms indicates the stress per atom in the x-direction (in eV).}
\label{figure_parallel_5nm_centre_first_second_pass_pinning_force}
\end{figure}

In Figure~\ref{figure_parallel_5nm_centre_first_second_pass_pinning_force} (a-d), we present the interaction between the edge dislocation and a larger, 5 nm vacancy loop. The 5 nm loop in its stable configuration is a closed loop, in line with the findings of \cite{gilbert2008structure}. The observations stated here hold true for vacancy loops in the size range 2-9 nm, in the parallel configuration. The interaction mechanism is clearly different from that observed for the 1 nm loop. On contact with the dislocation, the upper half of the loop is transformed into a vacancy cluster (Figure~\ref{figure_parallel_5nm_centre_first_second_pass_pinning_force} (c)) and the lower half of the loop is absorbed by the dislocation as a superjog. 

\begin{figure}[H]
\centering
\includegraphics[width = 0.8\linewidth]{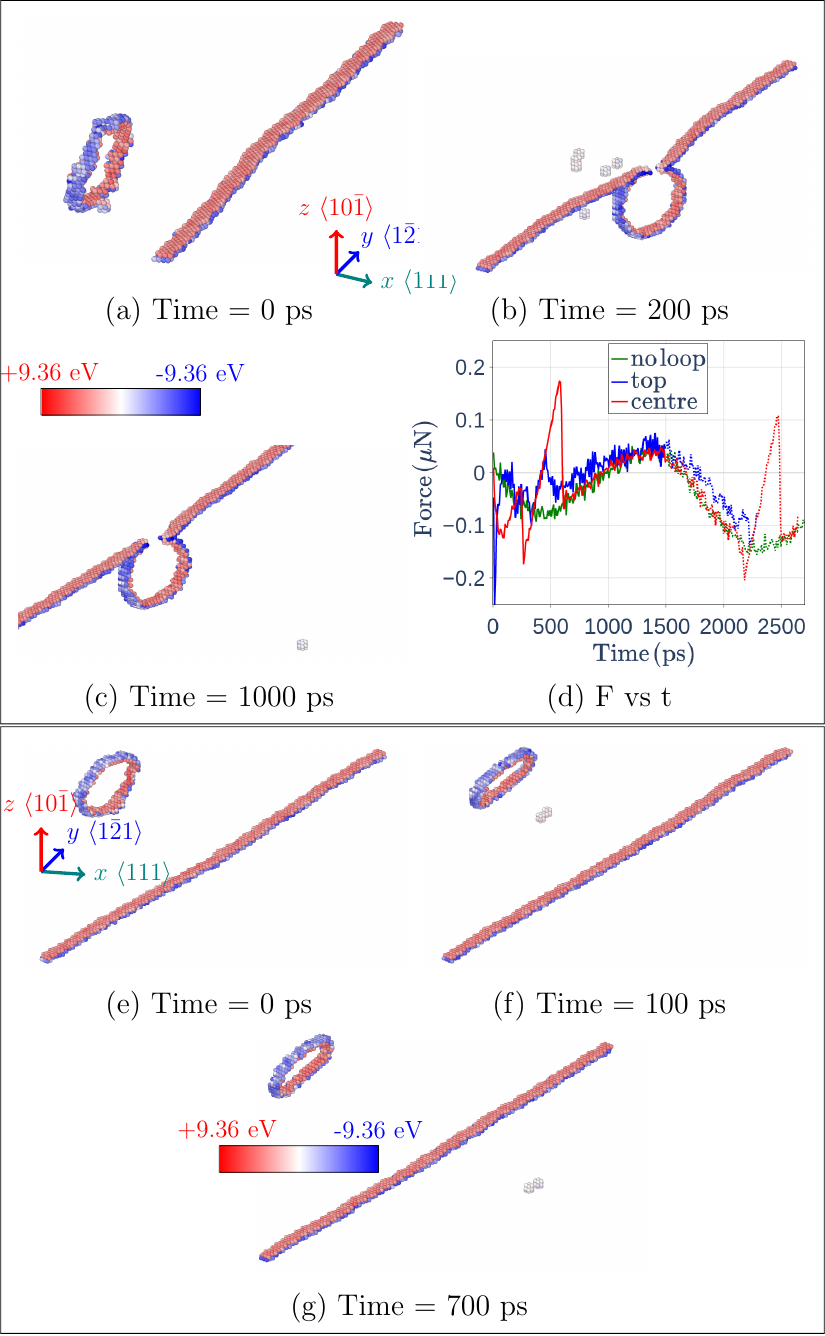}
\caption{Different stages of the interaction of an $\frac{1}{2}\langle 111\rangle$ edge dislocation with a $\frac{1}{2}\langle 111\rangle$ vacancy loop of size 5 nm, when the dislocation meets the loop at the top ((a)-(c)) and bottom ((e)-(g)), during the first pass of the dislocation through the simulation cell (interaction type A, see Table~\ref{table_loop_dislocation_interaction}. The corresponding curve of force \textit{vs.} time is shown in (d). The colouring of the atoms indicates the stress per atom in the x-direction (in eV).}
\label{figure_parallel_top_bottom_d_5nm_first_pass_pinning_force}
\end{figure}

The corresponding curve of pinning force vs time is shown in (solid) red in Figure~\ref{figure_parallel_top_bottom_d_5nm_first_pass_pinning_force} (d). Here, the peak corresponds to the unpinning of the dislocation with the superjog from the remnant vacancy cluster. The unpinning process involves leaving behind more vacancies in its wake, as it moves forward, as compared to the 1 nm loop (see Figure~\ref{figure_parallel_centre_d_1nm_top_centre_first_pass_pinning_force} (d)).

\begin{flushleft}
\textbf{\small Interaction D}
\end{flushleft}

Next, we consider the interaction between the remnant vacancy defect (remaining from the 5 nm vacancy loop after Interaction A) and the unpinned dislocation with the superjog. Because of the periodic boundary condition imposed along the X-direction, the unpinned dislocation exits the simulation cell on one side and reappears from the opposite side to encounter the remnant vacancy loop. The corresponding pinning force \textit{vs.} time curve is shown as a dotted red curve in Figure~\ref{figure_parallel_top_bottom_d_5nm_first_pass_pinning_force} (d). 
Unlike the initial interaction, where the dislocation interacts with an intact prismatic vacancy loop (Interaction A), the second interaction occurs with a remnant vacancy cluster generated during the first dislocation interaction (Interaction D). Consequently, the obstacle encountered by the returning dislocation is different in both morphology and size. This results in a significantly weaker obstacle, with the peak pinning strength decreasing by approximately 39 \% compared with the initial interaction. This highlights that obstacle strength evolves with defect evolution during plastic deformation.

These observations demonstrate that although larger parallel vacancy loops initially provide some resistance to dislocation motion, the interaction itself irreversibly transforms the defect into a substantially weaker obstacle. Consequently, the hardening contribution of these loops depends not only on their initial pinning strength but also on their configuration under repeated dislocation interaction.

\begin{flushleft}
\textit{Dislocation intersects top of loop}
\end{flushleft}

Figure~\ref{figure_parallel_top_bottom_d_5nm_first_pass_pinning_force} (a)-(c) shows the interaction mechanism when the dislocation intersects the top of the 5 nm loop. Unlike the case when the dislocation meets the centre of the loop, the top intersection results in no prominent pining strength (blue curve in Figure~\ref{figure_parallel_top_bottom_d_5nm_first_pass_pinning_force} (d)). The vacancy loop is attracted towards the dislocation, glides towards it (along the glide cylinder), and by $t = 200$ ps, the entire vacancy loop is absorbed by the gliding dislocation, producing a super-jogged dislocation without leaving behind a remnant vacancy trail or residual defect structure. Consequently, the original obstacle is completely eliminated during the first dislocation interaction through the simulation box. This modified dislocation structure passes through the cell for the second time, without any additional resistance, as shown by the dotted blue curve in Figure~\ref{figure_parallel_top_bottom_d_5nm_first_pass_pinning_force} (d). As expected, the pinning force \textit{vs.} time curve (in blue) for this superjogged dislocation has slightly higher resistance than the green curve (inherent lattice resistance experienced by the dislocation in a defect-free cell).

\begin{flushleft}
\textit{Dislocation intersects bottom of loop}
\end{flushleft}

In contrast to when the dislocation meets the centre and top of the loop, the bottom intersection does not result in a dislocation-loop intersection. As the dislocation approaches the loop, the loop moves away from the dislocation due to repulsive interaction Figure~\ref{figure_parallel_top_bottom_d_5nm_first_pass_pinning_force} (f). This continues during the second pass of the dislocation through the cell. In this configuration, the loop and dislocation glide together as a single unit. Consequently, even without intersecting, the dislocation and its elastic strain field move the loop from its initial position, leading to ``defect clearing''. Thus, despite the lack of direct intersection with the dislocation, the defect is still cleared.

These observations demonstrate that bottom intersection preserves the loop structure rather than eliminating it (as in the case of top intersection). However, both interactions lead to ``defect clearing''. Moreover, during centre interaction, the first dislocation pass through the cell modifies the loop configuration, leading to a comparatively weaker pinning of the dislocation during the second pass. So, the centre interaction represents a scenario in which repeated dislocation interactions progressively weaken the obstacle, rather than completely destroying/removing it (as observed during the top and bottom intersections). In other words, the intersection position governs the structural evolution and subsequent pinning strength of the vacancy loop following the dislocation interaction. The first interaction may either transform the loop into a weaker remnant defect, completely annihilate the obstacle by absorption, or carry the loop away without dislocation pinning. Consequently, the hardening contribution of a vacancy loop cannot be characterised solely by its initial obstacle strength but must also account for its persistence under repeated dislocation interactions. These findings imply that the evolution of irradiation hardening depends not only on the resistance offered by individual defects but also on how their obstacle strength changes as the defect landscape evolves during plastic deformation.

So far, we have presented the interaction mechanisms between edge dislocations and vacancy loops with parallel Burgers vectors; we have considered two loop sizes (1 nm and 5 nm, representing platelets and closed loops, respectively) and also shown the interaction mechanisms corresponding to three different intersection points for both loop sizes. For each considered interaction, the evolution of the interaction mechanism for the subsequent dislocation pass (Interaction D) and the resulting obstacle strengths are also shown. A summary of the interaction mechanisms in each of the above-explored cases is shown in Table~\ref{table_parallel_loop_meeting_position_interaction_mechanism_pinning_strengths}.

\begin{table}[H]
\centering
\begin{adjustbox}{width = \linewidth}
\begin{tabular}{|c|c|c|c|c|}
\hline
d    & Point of     & \multicolumn{2}{c|}{Interaction Mechanism and Pinning Strength} & Strength     \\
(nm) & intersection & \multicolumn{2}{c|}{}  & decline (\%) \\

\cline{3-4}
 & & First Pass  & Second Pass & \\
\hline
\multirow{3}{*}{1} 
  & Top    
  & \makecell{No pinning}
  & \makecell{No pinning}
  & - \\
\cline{2-5}
  & Centre 
  & \makecell{Interaction I \\ (Vacancy platelet) \\ (0.35$\mu$N)} 
  & \makecell{No (prominent) \\ pinning (0.05$\mu$N)}
  & 85.7 \\
\cline{2-5}
  & Bottom 
  & \makecell{Interaction I \\ (Vacancy platelet) \\ (0.1$\mu$N)}
  & \makecell{No (prominent) \\ pinning (0.05$\mu$N)}
  & 50 \\
\hline
\multirow{3}{*}{5} 
  & Top    
  & \makecell{Entire loop absorbed \\ (No pinning)} 
  & \makecell{No pinning}
  & - \\
\cline{2-5}
  & Centre 
  & \makecell{Bottom-half loop absorbed \\ as superjog (0.18$\mu$N)}
  & \makecell{(0.11$\mu$N)}
  & 39 \\
\cline{2-5}
  & Bottom 
  & \makecell{No intersection}
  & \makecell{No intersection}
  & - \\
\hline
\end{tabular}
\end{adjustbox}
\caption{Interaction mechanisms and pinning strengths during first and second pass as a function of loop-dislocation intersection position for parallel loops of sizes 1 nm and 5 nm; the percentage decline in pinning strength during second pass is listed in the last column. Interaction I means that the dislocation is pinned by the loop, it bows and breaks away, see Table~\ref{table_loop_dislocation_interaction_mechanism}}
\label{table_parallel_loop_meeting_position_interaction_mechanism_pinning_strengths}
\end{table}

\subsubsection{Discussion}

\begin{flushleft}
\textbf{\small Comparison with interstitial loop}
\end{flushleft}

The interaction mechanisms identified for parallel vacancy loops when the dislocation meets the loop centre can be compared with previous atomistic studies of edge dislocation interactions with interstitial loops. The interaction observed for the 1 nm vacancy platelet (``Interaction I'', see Table~\ref{table_loop_dislocation_interaction_mechanism} differs markedly from that reported for a similarly sized $\langle111\rangle$ interstitial loop interacting with a $\frac{1}{2}\langle111\rangle$ edge dislocation in pure Fe, where the interstitial loop is dragged by the gliding dislocation rather than being destroyed \cite{osetsky2004dynamic}. In contrast, the interaction mechanism identified for the 5 nm vacancy loop closely resembles that reported for similarly sized interstitial loops, where the defect is partially absorbed by the edge dislocation, leading to the formation of a superjog \cite{terentyev2013radiation, grammatikopoulos2019simulation}. However, the post-interaction evolution of the defect differs significantly. Following interaction with the dislocation, the remnant vacancy defect remains as a compact vacancy cluster (Figure~\ref{figure_parallel_5nm_centre_first_second_pass_pinning_force}(d)), whereas the remnant interstitial cluster spontaneously reconfigures into an interstitial loop \cite{terentyev2013radiation, yu2024atomistic}. These contrasting post-interaction modifications highlight that the long-term evolution of irradiation-induced defects depends not only on the initial dislocation–defect interaction mechanism but also on the reconfiguration of the remnant defect, which ultimately governs the obstacle strength for subsequent dislocations.

\begin{flushleft}
\textbf{\small Evolution of obstacle strength due to repeated dislocation interactions}
\end{flushleft}

The present results demonstrate that the strengthening contribution of irradiation-induced vacancy loops depends not only on their initial obstacle strength but also on their structural evolution following dislocation interaction. Three different evolution pathways are identified: complete obstacle annihilation (exemplified by the 5 nm loop interacting at the loop top), partial structural transformation into a weaker remnant defect (seen when the dislocation interacts with the centre of the 5 nm loop), and persistence of the loop accompanied by transport with the gliding dislocation (seen for the dislocation intersecting at the bottom of 5 nm loop). The pathway is governed by the intersection position and determines whether the obstacle strength is eliminated, progressively reduced, or preserved during subsequent dislocation interactions.

Existing constitutive models generally describe irradiation softening using a single scalar evolution law (e.g., obstacle strength decreases with accumulated slip or dislocation density) \cite{shi2022experiments,das2020modified,lin2024atomic,li2021temperature,hu2016irradiation}. However, the present results show that both the structural evolution of irradiation-induced defects and the corresponding evolution of their obstacle strength depend on the interaction mechanism, which in turn is governed by the dislocation–defect intersection position. Consequently, irradiation hardening is a complex function of the initial strength or density of irradiation defects, as well as of how they evolve and retain their pinning strength during repeated dislocation interactions. How these geometry-dependent evolution pathways can be systematically incorporated into continuum constitutive models remains an important open question.

\begin{flushleft}
\textbf{\small Hypothesis for interpretation of intersection position dependence}
\end{flushleft}

The contrasting behaviour observed for a dislocation intersecting the top, centre and bottom of the loop suggests that the dislocation–loop interaction may proceed through two distinct stages. Prior to direct contact, the approach of the dislocation appears to be influenced by the long-range elastic interaction between the edge dislocation and the vacancy loop. The top interactions appear consistent with an attractive interaction, allowing the dislocation to approach the loop and subsequently absorb (for 5 nm loop) or pass through it (seen for 1 nm platelet) with little resistance. In contrast, the bottom interactions are consistent with a repulsive interaction that tends to maintain a finite separation between the dislocation and the loop. The influence of this repulsion appears to depend on loop size: for the 5 nm loop, the repulsion is sufficiently strong that direct cutting is largely avoided and the loop is transported ahead of the dislocation, whereas the smaller 1 nm platelet still undergoes contact and therefore provides a finite, albeit weaker, pinning strength.

However, once direct contact occurs, the obstacle strength appears to be governed primarily by the position of intersection rather than by the preceding elastic interaction. In particular, centre intersections allow the dislocation to interact with the largest cross-section of the loop, producing the highest pinning strengths for both the 1 nm platelet and the 5 nm loop. These observations therefore suggest that long-range elastic interactions may control whether and how the dislocation reaches the defect, whereas the dislocation–loop intersection position may determine the interaction mechanism and the associated structural evolution of the defect.

Thus, overall, the size and position of the interaction both show notable differences in the pinning behaviour, defect pinning strengths, and the resulting evolution of defect configurations and material strength for parallel-oriented loops. Now we investigate the effect of the loop orientation on the interaction mechanism with the edge dislocation, by considering inclined loops in the following section. 

\subsection{Inclined loop}
\label{section_inclined}

Having established the governing interaction mechanisms for parallel vacancy loops, we now investigate the role of loop orientation with respect to the dislocation. Vacancy loops with a Burgers vector inclined by $70.53^\circ$ to that of the gliding edge dislocation are considered to examine how changes in the orientation relationship between the defect and the dislocation influence the interaction mechanism and the structure of the remnant defect during repeated dislocation interactions.

\subsubsection{Dislocation interaction with vacancy platelets (open loops)}

In the inclined orientation, the platelet configuration is observed for loop sizes in the range 1-3 nm. However, the interaction mechanism for the 1 nm-sized defect differs from that of the 2-3 nm-sized defects and each is discussed in turn below. 

\begin{flushleft}
\textit{1 nm platelets}
\end{flushleft}

For a 1 nm platelet in an inclined orientation, the interaction mechanisms with the dislocation remain identical to those observed in the parallel orientation, during both passes through the cell (Interaction A and Interaction D). 

In the case of centre interaction, during the first pass (interaction A), although the interaction mechanism remains unchanged, the corresponding pinning strength is reduced to 0.15 $\mu$N, compared to 0.35 $\mu$N obtained in parallel orientation (Figure~\ref{figure_parallel_centre_d_1nm_top_centre_first_pass_pinning_force}). This may be due to the dislocation experiencing a larger interaction cross-sectional area with the platelet in the parallel orientation than in the inclined one. Moreover, for the bottom-top interaction, weak (0.08 $\mu$N) and no pinning are observed in the inclined orientation, respectively, which is consistent with the behaviour of the parallel 1 nm loop.

Thus, for the 1 nm platelet, the orientation relationship has little influence on the governing interaction mechanism, although a modest reduction in pinning strength is observed for the inclined orientation.

\begin{flushleft}
\textit{2 nm platelets}
\end{flushleft}

\begin{flushleft}
\textit{Dislocation intersects centre of platelet}
\end{flushleft}

\begin{flushleft}
\textbf{\small Interaction A}
\end{flushleft}

\begin{figure}[H]
\centering
\includegraphics[width = 0.8\linewidth]{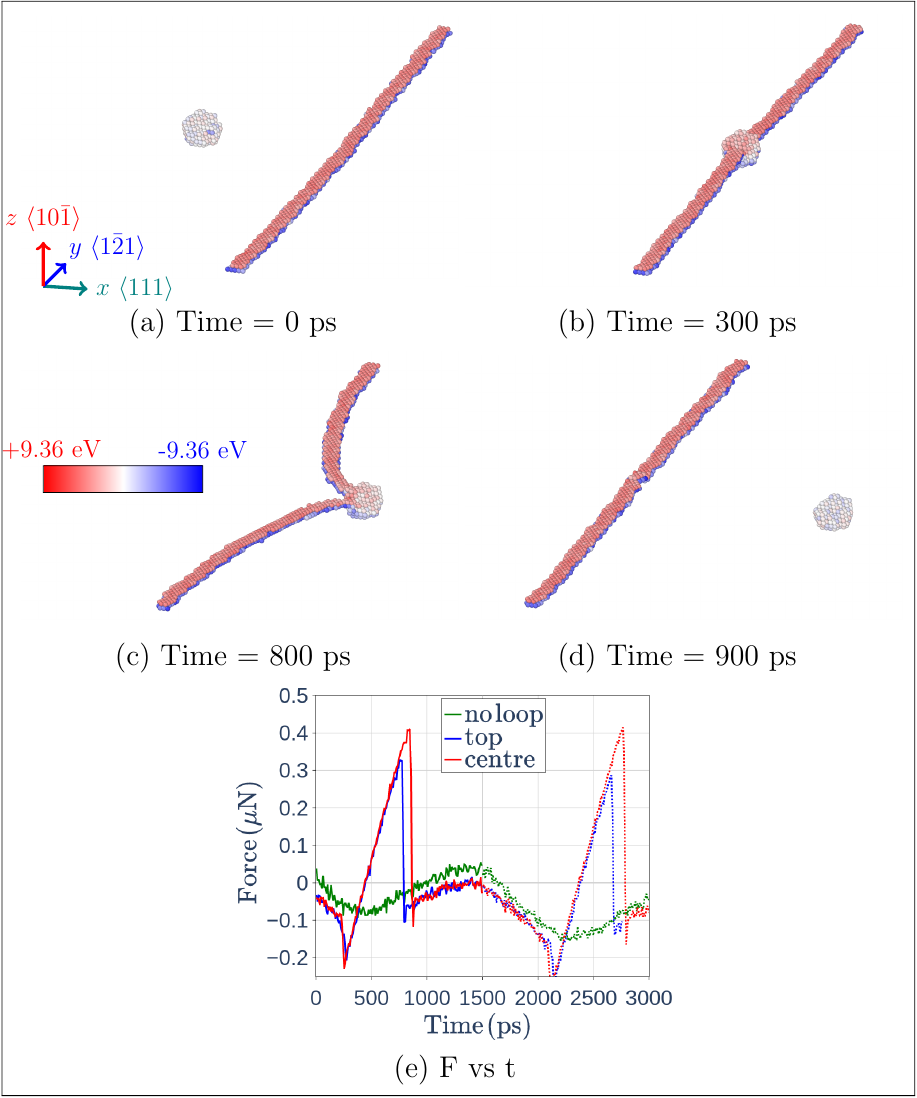}
\caption{Different stages of the interaction of an $\frac{1}{2}\langle 111\rangle$ edge dislocation with a $\frac{1}{2}\langle 1\bar11\rangle$ vacancy loop of size 2 nm, meeting at centre, during the first pass of the dislocation through the simulation cell (interaction type A, see Table~\ref{table_loop_dislocation_interaction}). The corresponding curve of pinning force \textit{vs.} time is shown in red in (e). The colouring of the atoms indicates the stress per atom in the x-direction (in eV).}
\label{figure_inclined_centre_2nm_first_pass_pinning_force}
\end{figure}

Figure~\ref{figure_inclined_centre_2nm_first_pass_pinning_force}(a)-(d) shows the different stages of the interaction of a $\frac{1}{2}\langle 111\rangle$ edge dislocation with a $\frac{1}{2}\langle 1\bar11\rangle$ vacancy platelet of size 2 nm. The interaction mechanism observed here is representative of those observed for loop sizes of 2-3 nm in the inclined orientation. An interaction mechanism similar to that defined as ``Interaction I'' occurs (see Table~\ref{table_loop_dislocation_interaction_mechanism}). The corresponding force \textit{vs.} time curve is shown in (solid) red in Figure~\ref{figure_inclined_centre_2nm_first_pass_pinning_force} (e). It is interesting to note that the lattice resistance experienced by the unpinned dislocation is lower than that for a pristine dislocation (green curve).

\begin{flushleft}
\textbf{\small Interaction D}
\end{flushleft}

The interaction of the unpinned dislocation with the remnant vacancy loop is shown in Figure~\ref{figure_appendix_inclined_centre_d_2nm_second_pass} in the Appendix~\ref{appendix_inclined_interaction_mechanism}.
As negligible vacancies are dislodged during Interaction A, the defect structure remains largely unaltered. Thus, the interaction mechanism and the pinning force in Interaction D, the second pass (dotted red curve in Figure~\ref{figure_inclined_centre_2nm_first_pass_pinning_force} (e)), are identical to the first pass.

\begin{flushleft}
\textit{Dislocation intersects top/bottom of platelet}
\end{flushleft}

\begin{flushleft}
\textbf{\small Interaction A}
\end{flushleft}

When the dislocation meets the top of the loop, an interaction mechanism, previously described as ``Interaction I'' (see Table~\ref{table_loop_dislocation_interaction_mechanism}), is observed. The corresponding pinning force \textit{vs.} time curve is shown in (solid) blue in Figure~\ref{figure_inclined_centre_2nm_first_pass_pinning_force} (f). The interaction mechanism at the top of the loop is nearly identical to that observed at the bottom of the loop.

\begin{flushleft}
\textbf{\small Interaction D}
\end{flushleft}

The interaction mechanism of the unpinned dislocation with the remnant vacancy loop remains ``Interaction I''; however, the pinning force is slightly lower (dotted blue curve in Figure~\ref{figure_inclined_centre_2nm_first_pass_pinning_force} (f)) because the remnant vacancy platelet left has fewer vacancies than the pristine version, owing to vacancies carried away by the dislocation during Interaction A.

These results demonstrate a size-dependent transition in the evolution of inclined vacancy platelets. Whereas the 1 nm platelet loses its obstacle strength following the initial interaction, the larger 2–3 nm platelets undergo negligible structural evolution and therefore retain both the interaction mechanism and obstacle strength during repeated dislocation interactions.

\subsubsection{Dislocation interaction with closed loops}

\begin{flushleft}
\textit{Dislocation intersects centre of loop}
\end{flushleft}

\begin{flushleft}
\textbf{\small Interaction A}
\end{flushleft}

\begin{figure}[H]
\centering
\includegraphics[width = 0.8\linewidth]{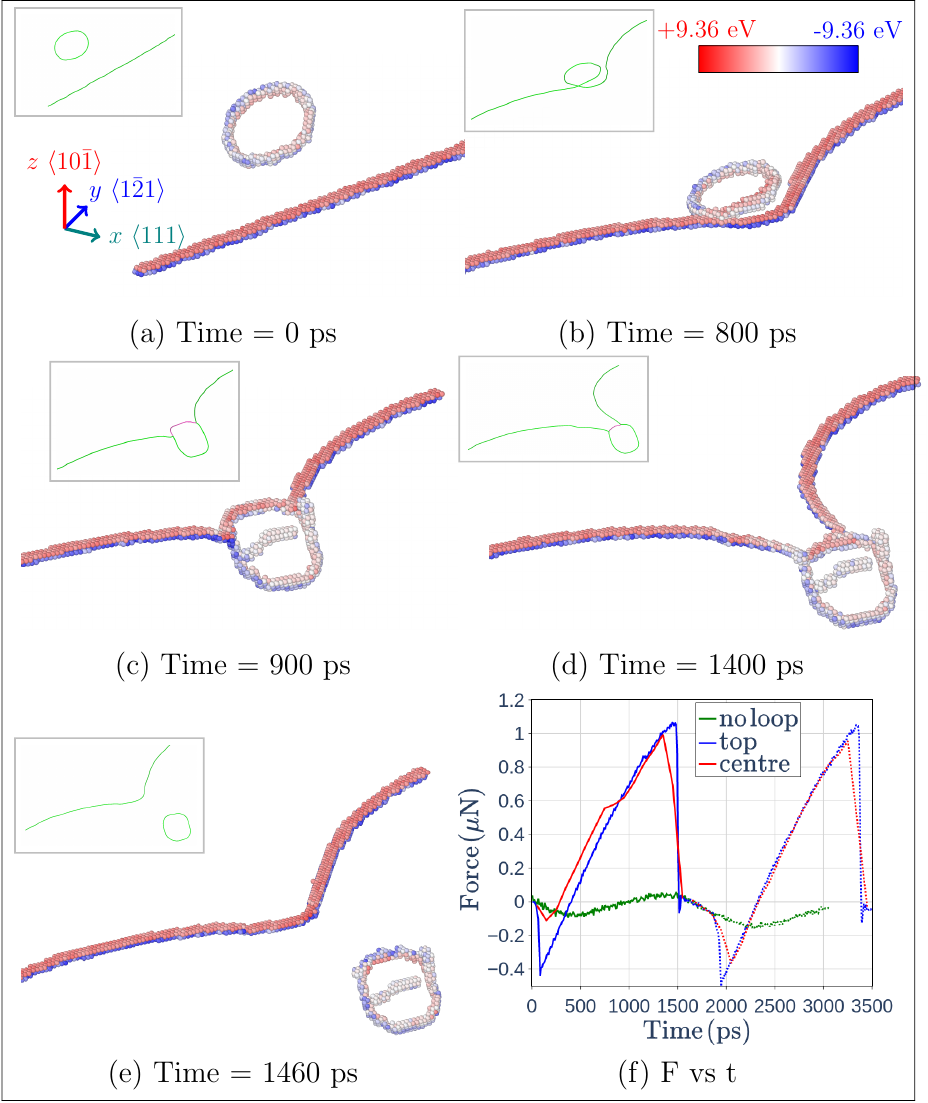}
\caption{Different stages of the interaction of an $\frac{1}{2}\langle 111\rangle$ edge dislocation with a $\frac{1}{2}\langle 1\bar11\rangle$ vacancy loop of size 5 nm, meeting at centre, during the first pass of the dislocation through the simulation cell (interaction type A, see Table~\ref{table_loop_dislocation_interaction}). The corresponding curve of force \textit{vs.} time is shown in red in (f). The insets show the Burgers vectors of the dislocations: $\frac{1}{2}\langle 111\rangle$ type in green and $\langle 010\rangle$ type in violet. The colouring of the atoms indicates the stress per atom in the x-direction (in eV).}
\label{figure_inclined_centre_d_5nm_first_pass}
\end{figure}

Figure~\ref{figure_inclined_centre_d_5nm_first_pass} (a-e) shows the different stages of the interaction of a $\frac{1}{2}\langle 111\rangle$ edge dislocation with a $\frac{1}{2}\langle 1\bar11\rangle$ vacancy loop of size 5 nm. The loop reorients as the dislocation approaches(Figure~\ref{figure_inclined_centre_d_5nm_first_pass} (b)). The repulsive interaction leads to bowing out of the dislocation. As stress increases, this repulsion is overcome and the dislocation meets the loops. On contact with the dislocation, the upper part of the loop is transformed into a $\frac{1}{2}\langle010\rangle$ segment (Figure~\ref{figure_inclined_centre_d_5nm_first_pass} (c)) by the following reaction:

\begin{equation}
    \frac{1}{2}\langle1 1 1\rangle \thickspace - \thickspace \frac{1}{2}\langle1 \bar1 1\rangle \thickspace \rightarrow \thickspace \langle010\rangle
    \label{equation_inclined_reaction}
\end{equation}

The insets in Figure~\ref{figure_inclined_centre_d_5nm_first_pass} show the type of dislocations formed, $\frac{1}{2}\langle 111\rangle$ type in green and $\langle010\rangle$ type in violet. After the dislocation reaction, the lower half of the loop still retains the original Burgers vector $\frac{1}{2}\langle 1\bar11\rangle$. Because the $\langle010\rangle$ segment is sessile on the $\langle10\bar1\rangle$ slip plane, it acts as a strong pinning point for the remaining segments of the $\frac{1}{2}\langle 111\rangle$ dislocation. With time, stress increases and these segments bow out into screw orientation (Figure~\ref{figure_inclined_centre_d_5nm_first_pass} (c) and (d)), forming a screw dipole. The screw segments then move towards each other by cross-slip, and annihilate, thereby unpinning the dislocation from the loop (Figure~\ref{figure_inclined_centre_d_5nm_first_pass} (e)). This mechanism of interaction, involving the formation of the sessile $\langle010\rangle$ segments, will be referred to hereon as ``Interaction II'' (see Table~\ref{table_loop_dislocation_interaction_mechanism}). The remnant loop is a distorted closed-form structure with a few vacancies dislodged by the unpinning dislocation. The corresponding pinning force \textit{vs.} time curve is shown in solid red in Figure~\ref{figure_inclined_centre_d_5nm_first_pass} (f).

\begin{flushleft}
\textbf{\small Interaction D}
\end{flushleft}

In this case, i.e., the second pass of the unpinned dislocation through the cell, where it meets the ``remnant defect'', the interaction mechanism is identical to that observed in Interaction A. The vacancy loop again forms $\langle010\rangle$ segment by reacting with the approaching dislocation, followed by bowing of remaining segments of the dislocation into screw orientation, subsequent cross-slip and unpinning. No measurable reduction in the peak pinning strength is observed during the second interaction (dotted red curve in Figure~\ref{figure_inclined_centre_d_5nm_first_pass} (f)), indicating that the initial dislocation interaction does not significantly alter the obstacle presented by the inclined vacancy loop.

\begin{flushleft}
\textit{Dislocation intersects top/bottom of loop}
\end{flushleft}

The top and bottom interactions exhibit behaviour analogous to that observed when the dislocation intersects the loop centre. The interaction mechanism, involving loop reorientation followed by repulsive interaction and dislocation breaking through, remains unchanged during the second dislocation interaction. Likewise, no measurable reduction in the peak pinning strength is observed (blue curve in Figure~\ref{figure_inclined_centre_d_5nm_first_pass} (f)), indicating that this intersection position also preserves the obstacle strength following repeated dislocation interactions.

Overall, these observations demonstrate that, unlike the parallel vacancy loops, the strengthening contribution of inclined vacancy loops is essentially unaffected by repeated dislocation interactions. Regardless of the intersection position, the governing interaction mechanism is preserved during successive dislocation interactions. Consequently, both the pinning strength and obstacle persistence remain largely unchanged, indicating that inclined vacancy loops represent highly stable irradiation defects capable of sustaining repeated dislocation interactions.

\subsubsection{Discussion}

Table~\ref{table_inclined_loop_meeting_position_interaction_mechanism_pinning_strengths} lists the interaction mechanisms, pinning strengths, and percentage decline in pinning strength during the first and second passes as a function of loop-dislocation intersection position for inclined loops of sizes 2 nm and 5 nm.

\begin{flushleft}
\textbf{\small Comparison with interstitial loop}
\end{flushleft}

The interaction mechanisms of larger inclined vacancy loops (5 - 10 nm) with edge dislocations are similar to those previously reported in the case of interstitial loops \cite{yu2024atomistic}. This corresponds to ``Interaction II'' (see Table~\ref{table_loop_dislocation_interaction_mechanism}), in which the $\langle 100\rangle$ segment is formed, the remaining $\langle 111\rangle$ segments bow out into a screw orientation, and the dislocation subsequently unpins following cross slip. 

In contrast, the interaction behaviour observed for smaller vacancy platelets (1–3 nm) differs markedly from that previously reported for a $\langle111\rangle$ interstitial loop of comparable size. In case of vacancy loops, the interaction mechanism corresponds to ``Interaction I'' (see Table~\ref{table_loop_dislocation_interaction_mechanism}), with the energetically favoured loop configuration being the vacancy platelet. However, for interstitial loops of similar size, the stable structure is a closed loop and therefore, a different interaction mechanism is observed, where $\langle 100 \rangle$ segments are formed.

\begin{table}[H]
\centering
\begin{adjustbox}{width = \linewidth}
\begin{tabular}{|c|c|c|c|c|}
\hline
d    & Point of     & \multicolumn{2}{c|}{Interaction Mechanism and Pinning Strength} & Strength     \\
(nm) & intersection & \multicolumn{2}{c|}{}  & decline (\%) \\
\cline{3-4}
 & & First Pass  & Second Pass & \\
\hline
\multirow{3}{*}{2} 
  & Top    
  & \makecell{Interaction I \\ (Vacancy platelet) \\ (0.33 $\mu$N)}
  & \makecell{Interaction I \\ (Vacancy platelet) \\ (0.29 $\mu$N)}
  & 12 \\
\cline{2-5}
  & Centre 
  & \makecell{Interaction I \\ (Vacancy platelet) \\ (0.42 $\mu$N)} 
  & \makecell{Interaction I \\ (Vacancy platelet) \\ (0.42 $\mu$N)}
  & 0 \\
\cline{2-5}
  & Bottom 
  & \makecell{Interaction I \\ (Vacancy platelet) \\ (0.33 $\mu$N)}
  & \makecell{Interaction I \\ (Vacancy platelet) \\ (0.26 $\mu$N)}
  & 12 \\
\hline
\multirow{3}{*}{5} 
  & Top    
  & \makecell{Interaction II \\ (1.07$\mu$N)} 
  & \makecell{Interaction II \\ (Remnant closed loop) \\ (1.07$\mu$N)}
  & 0.9 \\
\cline{2-5}
  & Centre 
  & \makecell{Interaction II \\ (1.05$\mu$N)}
  & \makecell{Interaction II \\ (Remnant closed loop) \\ (0.99$\mu$N)}
  & 4.8 \\
\cline{2-5}
  & Bottom 
  & \makecell{Interaction II \\ (1.1$\mu$N)}
  & \makecell{Interaction II \\ (Remnant closed loop) \\ (1.04$\mu$N)}
  & 6.25 \\
\hline
\end{tabular}
\end{adjustbox}
\caption{Interaction mechanisms and pinning strengths during first and second pass as a function of loop-dislocation intersection position for inclined loops of sizes 1 nm and 5 nm; the percentage decline in pinning strength during second pass is listed in the last column. ``Interaction I'' means that the dislocation is pinned by the loop, it bows and breaks away and ``Interaction II'' means $\langle010\rangle$ segment forms, which pins the gliding dislocation; the free segments bow out and break away, refer to Table~\ref{table_loop_dislocation_interaction_mechanism}}
\label{table_inclined_loop_meeting_position_interaction_mechanism_pinning_strengths}
\end{table}

\begin{flushleft}
\textbf{\small Comparison with parallel orientation}
\end{flushleft}

We note that the vacancy platelets are the stable configuration only up to approximately 1 nm in the parallel orientation, but up to about 3 nm in the inclined orientation.

Analysing all the mechanisms explored across the parallel and inclined orientations, it can be concluded that the platelet form (or open form) of the vacancy loop always exhibits ``Interaction I'' (see Table~\ref{table_loop_dislocation_interaction_mechanism}) while meeting the dislocation. But the strength of the platelet is higher when it meets the dislocation in parallel orientation. This is seen by comparing the dislocation interactions of the 1 nm loop in parallel (0.34 $\mu$N) \textit{vs.} inclined orientations (0.15 $\mu$N), both of which exist as vacancy platelets during interaction. This may be due to the dislocation experiencing a larger interaction cross-sectional area with the platelet in the parallel orientation than in the inclined one.

The pinning strength of the closed loops is higher in the inclined configuration due to the formation of the sessile $\langle010\rangle$ dislocation segments during dislocation interaction with the inclined closed loops. This is clearly illustrated by considering the case where the dislocation meets the centre of the 5 nm (closed) loop. The strength in inclined orientation is $\sim$6 times higher than the corresponding strength in the parallel orientation.

\begin{flushleft}
\textbf{\small Influence of size on interaction mechanism}
\end{flushleft}

The present results demonstrate that increasing defect size influences not only the obstacle strength but also the underlying interaction mechanism. For inclined vacancy platelets, increasing the platelet size from 1 nm to 2–3 nm results in a transition from defects that undergo structural transformation during the first dislocation interaction to defects that remain largely intact following repeated interactions. Consequently, the obstacle persists during multiple dislocation interactions beyond 2 nm. A further increase in defect size into the closed-loop regime produces a different interaction mechanism involving dislocation reactions, sessile $\langle100\rangle$ dislocation segments and screw-mediated unpinning. Within this size regime, larger loops provide progressively stronger pinning owing to the formation of longer sessile segments, as reflected by the increase in pinning strength from 1.05 $\mu$N for the 5 nm loop to 1.41 $\mu$N for the 8 nm loop. These observations indicate that increasing defect size influences both the governing interaction mechanism and the resulting strengthening behaviour.

\begin{flushleft}
\textbf{\small Evolution of obstacle strength due to repeated dislocation interactions}
\end{flushleft}

Repeated dislocation interactions reveal a clear transition in obstacle persistence with increasing defect size. The smallest 1 nm platelets lose their pinning strength following the initial interaction, irrespective of their orientation with respect to the dislocation, indicating that such small defects cannot sustain repeated interactions with gliding dislocations. In contrast, inclined platelets of 2–3 nm and all inclined closed loops retain both their interaction mechanism and obstacle strength during successive dislocation interaction because the remnant defects undergo negligible structural evolution. These observations suggest that the persistence of irradiation defects is primarily governed by the interaction mechanism (determined by defect orientation), which dictates the extent of structural evolution during dislocation interaction, rather than by the initial obstacle strength alone. 

\begin{flushleft}
\textbf{\small Comparison across intersection positions}
\end{flushleft}

The behaviour of the inclined loops differs entirely from that of the parallel loops. The observations from the parallel loops suggest that long-range elastic interactions may influence the initial approach of the dislocation and, in some cases, even prevent direct intersection (as in the case of 5 nm parallel loop bottom intersection). However, the results for the inclined loop do not indicate a dominant role of the long-range elastic interaction in determining the interaction outcome for the inclined loops. Across the centre, top and bottom interactions, both the governing interaction mechanism and the obstacle strength remain essentially unchanged. A possible explanation is that, unlike the parallel loops, all three inclined intersections yield the same dislocation reaction, producing a sessile $\langle100\rangle$ junction regardless of the initial point of contact. Consequently, the interaction mechanism appears to be dictated primarily by the dislocation reaction rather than by the intersection position. 

\begin{flushleft}
\textbf{\small Summary of observations from parallel and inclined loops}
\end{flushleft}

Taken together, the present simulations demonstrate that dislocation–loop interactions lead to different pathways of defect evolution: obstacle annihilation, transformation into a defect of altered strength, or persistence with relatively little change. Parallel loops exhibit strong dependence on the intersection position, which may result in loop absorption, transformation into weaker remnant defects, or transport of the loop by the gliding dislocation. In contrast, inclined loops undergo a distinct dislocation reaction involving the formation of sessile $\langle 100 \rangle$ segments, resulting in substantially higher and comparatively geometry-insensitive obstacle strengths. Thus, a gliding dislocation does not simply overcome a pre-existing obstacle; it can modify, remove, or transform the defect itself, thereby changing the obstacle encountered by subsequent dislocations. 

Our observations from the MD simulations is consistent with recent in-situ TEM observations in iron by \cite{inoue2026direct}, who experimentally studied the interaction mechanisms between gliding $\frac{1}{2}\langle 111 \rangle$ screw dislocations and $\langle 100 \rangle$ interstitial loops: loop restoration, absorption, and transformation into $\frac{1}{2}\langle 111 \rangle$ loops. Notably, they also observed multiple interaction mechanisms for loops of comparable sizes, reinforcing that the initial defect size alone is insufficient to determine the subsequent dislocation–loop reaction. Although the dislocation and loop character differ from those considered here, these observations provide independent experimental evidence that irradiation defects can undergo mechanism-dependent evolution during their interaction with the dislocation, supporting the results of the present atomistic simulations. 

\subsection{Pinning force \textit{vs.} loop size}
\label{section_pinning_force_vs_loop_size}

\begin{figure}[H]
    \centering
    \includegraphics[width = 0.6\linewidth]{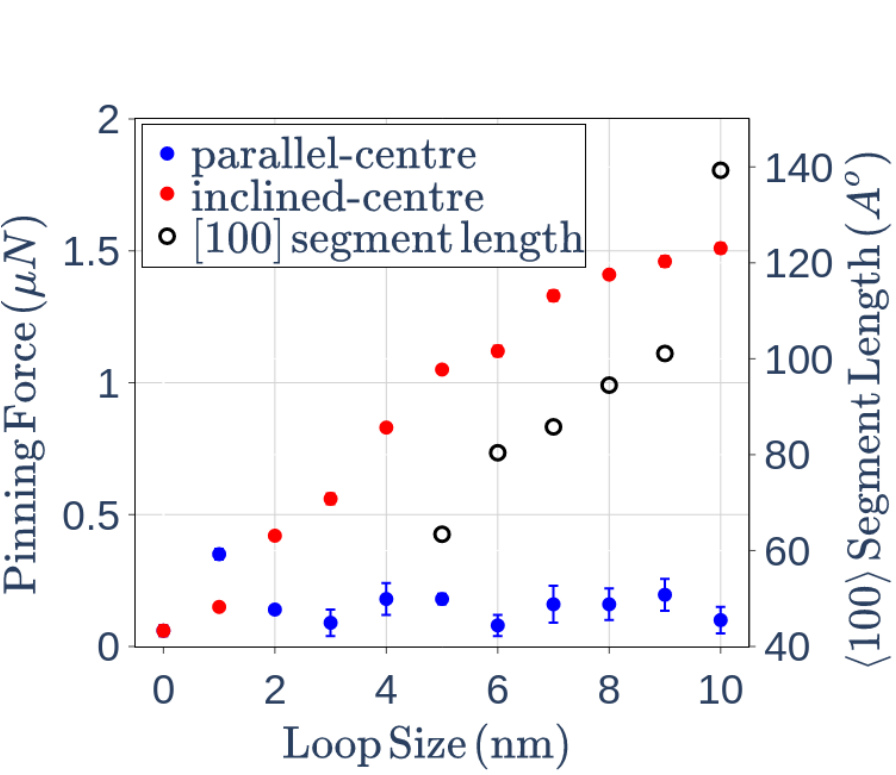}
\caption{Pinning force \textit{vs.} loop size for the interaction of an $\frac{1}{2}\langle 111\rangle$ edge dislocation with vacancy loop in parallel ($\frac{1}{2}\langle 111\rangle$) and inclined ($\frac{1}{2}\langle 1\bar11\rangle$) orientations. The error bar on the pinning force is calculated using averaging windows of lengths ranging from 1 ps to 20 ps, in steps of 1 ps.}
\label{figure_pinning_force_vs_size}
\end{figure}

Figure~\ref{figure_pinning_force_vs_size} shows the variation of pinning force with loop size for parallel and inclined vacancy loops when the dislocation intersects the loop centre. The error bar on the pinning force corresponds to the standard deviation calculated across a number of average values, where each average is calculated using averaging windows from 1 ps up to 20 ps, in steps of 1 ps. Here, two different size-dependent strengthening trends can be observed. For inclined loops, the pinning force increases monotonically with loop size, whereas for parallel loops, the pinning force is noticeably high only for a 1 nm sized loop, beyond which the pinning force decreases to an almost size-independent value for larger loops. These contrasting trends arise from the different interaction mechanisms governing the two loop orientations with respect to the gliding dislocation.

For the parallel orientation, vacancy platelets are the stable configuration only up to approximately 1 nm, beyond which closed loops become energetically favourable. Small platelets impede dislocation motion through ``Interaction I'' (the pinned dislocation bows out and eventually unpins, see Figures~\ref{figure_parallel_centre_d_1nm_top_centre_first_pass_pinning_force}), resulting in a modest obstacle strength. However, once the morphology transitions to a closed loop, the governing interaction mechanism changes. The vacancy loop is partially absorbed by the gliding dislocation, which reduces the effectiveness of the obstacle, resulting in the comparatively low and nearly size-independent pinning strength observed for larger loops, similar to that observed for interstitial loops in tungsten \cite{yu2024atomistic}.

The inclined orientation follows a markedly different size-dependent evolution. Vacancy platelets remain the stable configuration up to approximately 3 nm, during which the pinning strength increases progressively with platelet size as the larger defects are stronger obstacles to dislocation motion through ``Interaction I'' (see Table~\ref{table_loop_dislocation_interaction_mechanism}). Beyond this size, closed loops become the stable configuration, and the governing interaction mechanism changes, and the dislocation reacts with the loop to form sessile $\langle010\rangle$ dislocation segments (Interaction II, see Table~\ref{table_loop_dislocation_interaction_mechanism}), which strongly pin the gliding dislocation. As the loop size increases further, the length of the sessile $\langle010\rangle$ segment formed also increases, leading to a continued monotonic increase in pinning strength. This correlation is evident from the close agreement between the measured pinning force and the length of the sessile $\langle010\rangle$ segment shown in Figure~\ref{figure_pinning_force_vs_size}.

The nearly size-independent strengthening observed for larger parallel vacancy loops is consistent with previous atomistic studies of parallel interstitial loops, which also exhibit relatively weak, size-independent obstacle strengths \cite{yu2024atomistic}. In contrast, the monotonic increase in pinning strength observed for inclined vacancy loops follows the same qualitative trend reported for inclined interstitial loops \cite{yu2024atomistic}, although the magnitude of the strengthening differs owing to the different underlying defect structures and interaction mechanisms.

\begin{flushleft}
\textbf{\small Size-dependent evolution of remnant defect structures}
\end{flushleft}

The evolution of the remnant defect structure also depends strongly on the loop size. For parallel loops between 2 and 5 nm, the first dislocation interaction transforms the original loop into a vacancy cluster. In contrast, for larger loops (6–10 nm), the upper half of the loop remains as a loop following dislocation interaction. These observations indicate that continued plastic deformation progressively changes not only the obstacle strength but also the morphology of the irradiation defects, increasing the fraction of vacancy clusters generated from initially closed vacancy loops.

\subsection{Implications for irradiation hardening}

Previous work has demonstrated that crystal plasticity finite element (CPFE) modelling can reproduce the experimentally observed mechanical response of such low dose (0.01 dpa) self-ion irradiated tungsten by incorporating an effective irradiation-induced hardening contribution into the constitutive description \cite{das2020modified}. In that formulation, the interaction between gliding dislocations and vacancy loops (in a similar size range as investigated here) was represented using a simplified description based on the energetic cost of jog formation. The resulting effective irradiation-hardening contribution was calibrated against the experimental mechanical response of ion-implanted tungsten, and the CPFE model successfully reproduced the observed deformation behaviour \cite{das2020modified}.

The present molecular dynamics simulations show that the simplified jog-formation representation does not capture the full range of defect-evolution pathways. Rather than being governed by a single, fixed interaction mechanism, the simulations show that the response of a vacancy loop depends on its size, orientation, and the position at which the dislocation intersects the defect. Depending on these factors, the loop may persist, be annihilated, be transported, or transform into a remnant defect structure with a different obstacle strength. Furthermore, subsequent interactions with these remnant configurations can produce different responses from the initial interaction. The irradiation-induced defect population therefore evolves during plastic deformation, and the obstacle presented to a subsequent dislocation need not be equivalent to that presented by the original defect.

Taken together, these results highlight the value of atomistic simulations in providing a basis for interpreting and, in future, improving continuum descriptions of irradiation hardening. In particular, the present simulations show that the obstacle strength of vacancy loops cannot, in general, be characterised solely by the initial defect size, but may evolve as a consequence of multiple subsequent dislocation-loop interactions. 

\section{Conclusion}
\label{section_conclusion}

Molecular dynamics simulations were used to investigate edge dislocation-vacancy loop interactions in tungsten and to determine how successive interactions modify the irradiation-induced defect population. By varying loop size, orientation, and dislocation-loop intersection position, and by examining repeated interactions with remnant defects, we show that vacancy loops act as dynamic obstacles to dislocation motion, whose obstacle strength can evolve during plastic deformation. The key findings are as follows:

\begin{enumerate}

    \item Loop size and orientation determine the interaction mechanism and obstacle strength through their influence on loop morphology. At small sizes, parallel and inclined platelets exhibit similar behaviour, whereas at intermediate sizes, their stable morphologies become orientation-dependent. At larger sizes, parallel closed loops exhibit comparatively weak and nearly size-independent pinning, while inclined loops form sessile $\langle100\rangle$-type segments and show substantially stronger, size-dependent pinning.

    \item Intersection position affects the interaction mechanism and subsequent evolution of parallel loops, which may be absorbed, transformed, or transported with the dislocation. Inclined loops are substantially less sensitive to intersection position and exhibit similar interaction mechanisms and obstacle strengths for different intersection locations.

    \item Repeated dislocation interaction modifies both the defect structure and its obstacle strength. Small platelets can lose their pinning ability after the first interaction, while larger inclined defects generally retain their obstacle strength. Parallel closed loops can undergo substantial transformation or absorption, resulting in reduced or eliminated pinning during subsequent interactions.

    \item Dislocation-loop interactions therefore produce distinct defect-evolution pathways, including annihilation, transformation, transport, and persistence. Consequently, the obstacle landscape encountered during continued deformation can differ from that defined by the initial defect population.

    \item Obstacle strength of vacancy loops is therefore governed not only by initial defect size, but also by defect morphology, orientation, intersection position, and prior interaction history. These results provide atomistic insight into the evolving obstacle landscape underlying irradiation hardening, that should be considered in physically informed descriptions of irradiated plasticity.

\end{enumerate}

\section{Author contributions}
\textbf{Soumya Mishra}: Formal analysis, Investigation, Validation, Visualisation, Writing – original draft.

\textbf{Suchandrima Das}: Conceptualisation, Funding acquisition, Methodology, Resources, Supervision, Validation, Writing – review \& editing.

\section{Declaration of competing interest}
The authors declare that they have no known competing financial interests or personal relationships that could have influenced the work reported in this paper.

\section{Declaration of generative AI use}
During the preparation of this work, the author(s) used Grammarly, TextGPT (Overleaf), ChatGPT (OpenAI) and Claude (Anthropic) for grammar correction, sentence restructuring and paraphrasing. After using this tool/service, the authors have reviewed and edited the content as needed and take full responsibility for the content of the published article.

\section{Acknowlegments}
We gratefully acknowledge the Indian Institute of Science (IISc) for providing the computational facilities and funding this work through the Faculty Start-up Grant.

\renewcommand{\thefigure}{\thesection\arabic{figure}}
\makeatletter
\@addtoreset{figure}{section}
\makeatother

\renewcommand{\thetable}{\thesubsection\arabic{table}}
\makeatletter
\@addtoreset{table}{subsection}
\makeatother

\appendix

\section{Strain rate dependence of (pinning) force}
\label{appendix_strain_rate_dependence}

\begin{figure}[H]
\centering
\includegraphics[width = 1.0\linewidth]{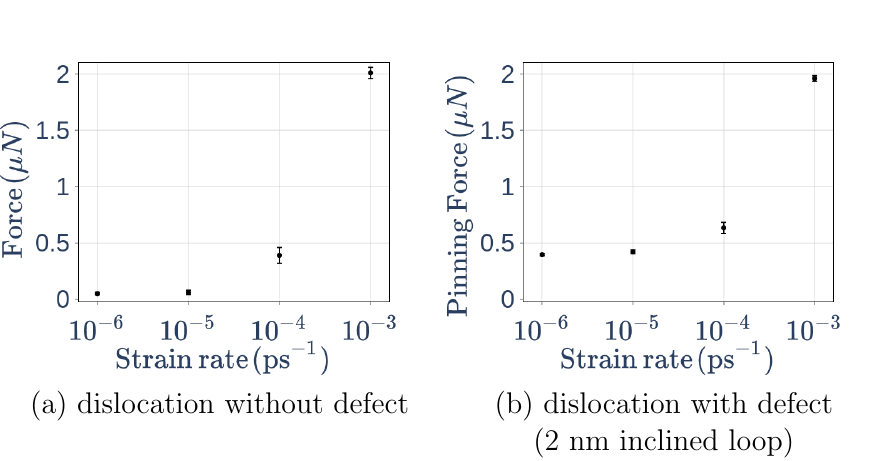}
\caption{(Pinning) force \textit{vs.} strain rate for an edge dislocation through the simulation cell, in (a) absence of any defect and (b) presence of inclined 2 nm loop. The error bar on the pinning force corresponds to the standard deviation calculated across a number of average values, where each average is calculated using averaging windows from 1 ps up to 20 ps, in steps of 1 ps.}
\label{figure_dislocation_F_vs_strain_rate}
\end{figure}

\section{Dislocation intersects 1 nm parallel vacancy platelet at centre}
\label{appendix_parallel_interaction_mechanism}

\begin{figure}[H]
\centering
\includegraphics[width = 0.8\linewidth]{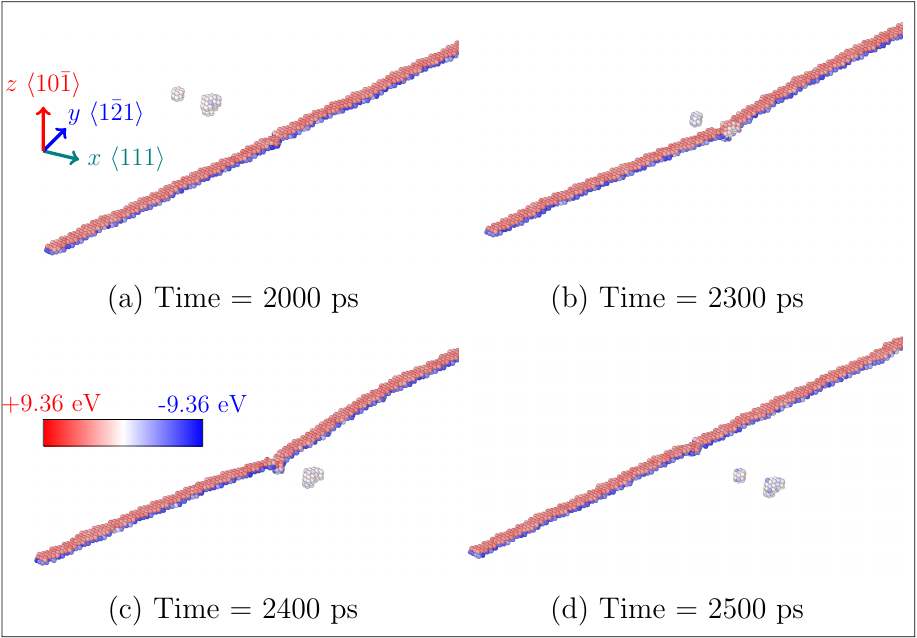}
\caption{Different stages of the interaction of an $\frac{1}{2}\langle 111\rangle$ edge dislocation with a $\frac{1}{2}\langle 111\rangle$ vacancy loop of size 1 nm, during the second pass of the dislocation through the simulation cell (interaction type D, see Table~\ref{table_loop_dislocation_interaction}). The corresponding curve of pinning force \textit{vs.} time is shown using the dotted red curve in Figure~\ref{figure_parallel_centre_d_1nm_top_centre_first_pass_pinning_force} (e). The colouring of the atoms indicates the stress per atom in the x-direction (in eV).}
\label{figure_appendix_parallel_centre_d_1nm_second_pass}
\end{figure}

\section{Dislocation intersects 2 nm inclined vacancy platelet at centre}
\label{appendix_inclined_interaction_mechanism}

\begin{figure}[H]
\centering
\includegraphics[width = 0.8\linewidth]{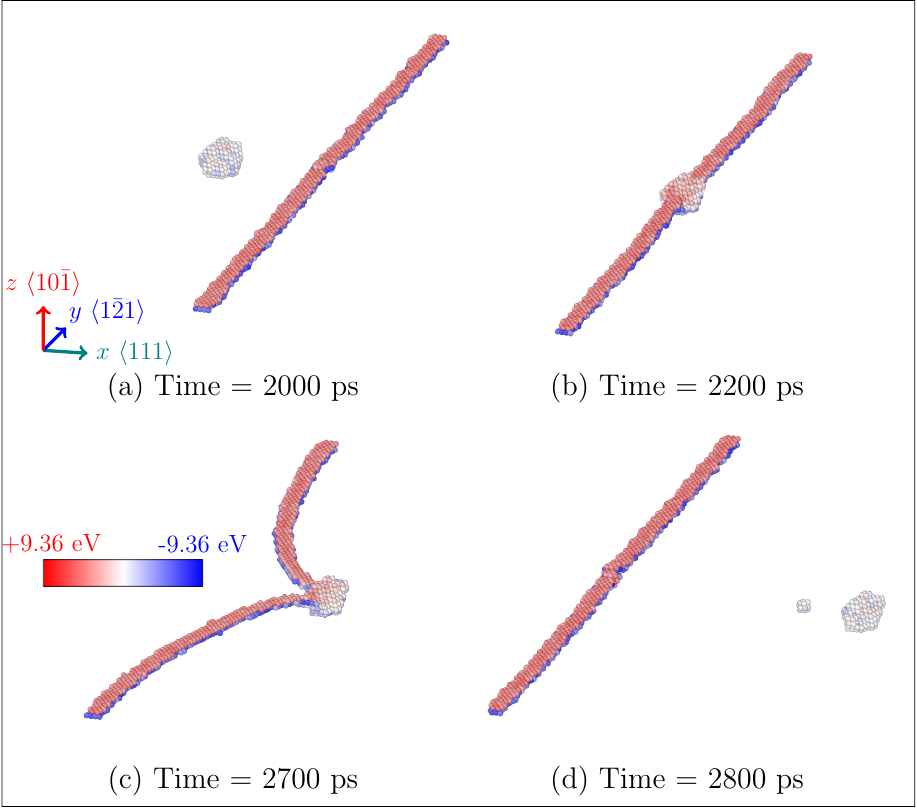}
\caption{Different stages of the interaction of an $\frac{1}{2}\langle 111\rangle$ edge dislocation with a $\frac{1}{2}\langle 1\bar11\rangle$ vacancy loop of size 2 nm, during the second pass of the dislocation through the simulation cell (interaction type D, see Table~\ref{table_loop_dislocation_interaction}). The corresponding curve of pinning force \textit{vs.} time is shown in red in Figure~\ref{figure_inclined_centre_2nm_first_pass_pinning_force} (e). The colouring of the atoms indicates the stress per atom in the x-direction (in eV).}
\label{figure_appendix_inclined_centre_d_2nm_second_pass}
\end{figure}

\section{Loop size and number of vacancies}
\label{appendix_loop_size_num_vacancies}

\begin{table}[H]
\centering
    \begin{tabular}{|c|c|c|}
    \hline
    \multirow{2}{*}{Loop size (nm)} & \multicolumn{2}{|c|}{Number of vacancies} \\
    \cline{2-3}   
     & Parallel & Inclined \\
    \hline
        1  & 14   & 12 \\
        2  & 57   & 53 \\
        3  & 121  & 121 \\ 
        4  & 215  & 219 \\
        5  & 339  & 337 \\
        6  & 491  & 489 \\
        7  & 663  & 666 \\
        8  & 863  & 870 \\
        9  & 1098 & 1099 \\
        10 & 1362 & 1356 \\
    \hline    
    \end{tabular}
    \caption{Number of vacancies in the parallel and inclined loops}
    \label{table_parallel_inclined_num_vacancies}
\end{table}

Table~\ref{table_parallel_inclined_num_vacancies} lists the diameter of the loops with parallel and inclined orientations and the corresponding number of vacancies in the pristine form.

\bibliographystyle{elsarticle-num}
\bibliography{references_loop_dislocation_tungsten}

@article{mason2014elastic,
year = {2014},
publisher = {IOP Publishing},
volume = {26},
number = {37},
pages = {375701},
author = {Mason, D R and Yi, X and Kirk, M A and Dudarev, S L},
title = {Elastic trapping of dislocation loops in cascades in ion-irradiated tungsten foils},
journal = {Journal of Physics: Condensed Matter},
}

@article{inoue2026direct,
  title={Direct in situ transmission electron microscopy observation of the interaction between gliding 1/2< 111> screw dislocations and< 100> prismatic dislocation loops in $\alpha$-iron},
  author={Inoue, T and Komatsu, M and Maekawa, T and Hatakeyama, T and Zhao, B and Kitano, M and Ueda, D and Sugimoto, Y and Dudarev, SL and Reed, RC and others},
  journal={Acta Materialia},
  pages={122067},
  year={2026},
  publisher={Elsevier}
}

@article{lin2024atomic,
  title={Atomic irradiation defects induced hardening model in irradiated tungsten based on molecular dynamics and CPFEM},
  author={Lin, Pan-dong and Nie, Jun-feng and Lu, Yu-peng and Shi, Chang-xin and Cui, Shu-gang and Cui, Wen-dong and He, Lei},
  journal={International Journal of Plasticity},
  volume={174},
  pages={103895},
  year={2024},
  publisher={Elsevier}
}

@article{ji2023atomistically,
  title={Atomistically-informed hardening and kinetics models of helium bubble in irradiated tungsten},
  author={Ji, Changqiu and Hu, Jianqiao and Zhuang, Zhuo and Cui, Yinan},
  journal={International Journal of Plasticity},
  volume={165},
  pages={103620},
  year={2023},
  publisher={Elsevier}
}

@article{xie2026atomistic,
  title={Atomistic insights into the enhancement of dynamic yield and fracture responses by chemical short-range order in CoCrNi multi-principal element alloy under quasi-isentropic loading},
  author={Xie, Zhuocheng and Jian, Wu-Rong and Xu, Shuozhi and Beyerlein, Irene J and Zhang, Xiaoqing and Zhang, Ningbo and Yao, Xiaohu},
  journal={Acta Materialia},
  volume={312},
  pages={122263},
  year={2026},
  publisher={Pergamon}
}

@article{shi2022experiments,
  title={Experiments and/or crystal plasticity finite element modeling of the mechanical properties of pristine and irradiated tungsten single crystal},
  author={Shi, Jiaqing and Liu, Guisen and Wu, Kaitao and Yu, Ping and Zhu, Heng and Zhao, Guannan and Shen, Yao},
  journal={International Journal of Plasticity},
  volume={154},
  pages={103293},
  year={2022},
  publisher={Elsevier}
}

@article{li2021temperature,
  title={Temperature dependent deformation localization in irradiated tungsten},
  author={Li, Zhijie and Liu, Zhanli and Zhuang, Zhuo and Cui, Yinan},
  journal={International Journal of Plasticity},
  volume={146},
  pages={103077},
  year={2021},
  publisher={Elsevier}
}

@article{hu2016irradiation,
  title={Irradiation hardening of pure tungsten exposed to neutron irradiation},
  author={Hu, Xunxiang and Koyanagi, Takaaki and Fukuda, Makoto and Kumar, NAP Kiran and Snead, Lance L and Wirth, Brian D and Katoh, Yutai},
  journal={Journal of Nuclear Materials},
  volume={480},
  pages={235--243},
  year={2016},
  publisher={Elsevier}
}

@article{song2024deformation,
  title={Deformation localisation in ion-irradiated Fe and Fe10Cr},
  author={Song, Kay and Sheyfer, Dina and Liu, Wenjun and Tischler, Jonathan Z and Das, Suchandrima and Mizohata, Kenichiro and Yu, Hongbing and Armstrong, David EJ and Hofmann, Felix},
  journal={Journal of Nuclear Materials},
  volume={596},
  pages={155104},
  year={2024},
  publisher={Elsevier}
}

@article{erinosho2015strain,
  title={Strain localization and failure in irradiated zircaloy with crystal plasticity},
  author={Erinosho, TO and Dunne, FPE},
  journal={International Journal of Plasticity},
  volume={71},
  pages={170--194},
  year={2015},
  publisher={Elsevier}
}

@article{zan2023nanoindentation,
  title={Nanoindentation study of $\delta$-phase zirconium hydride using the crystal plasticity model},
  author={Zan, XD and Guo, X and Weng, GJ and Chen, G},
  journal={International Journal of Plasticity},
  volume={167},
  pages={103675},
  year={2023},
  publisher={Elsevier}
}

@article{xiao2019modelling,
  title={Modelling nano-indentation of ion-irradiated FCC single crystals by strain-gradient crystal plasticity theory},
  author={Xiao, Xiazi and Chen, Lirong and Yu, Long and Duan, Huiling},
  journal={International Journal of Plasticity},
  volume={116},
  pages={216--231},
  year={2019},
  publisher={Elsevier}
}

@article{li2014predicting,
  title={Predicting plastic flow and irradiation hardening of iron single crystal with mechanism-based continuum dislocation dynamics},
  author={Li, Dongsheng and Zbib, Hussein and Sun, Xin and Khaleel, Mohammad},
  journal={International Journal of Plasticity},
  volume={52},
  pages={3--17},
  year={2014},
  publisher={Elsevier}
}

@article{ogorodnikova2013tem,
  title={TEM observations of radiation damage in tungsten irradiated by 20 MeV W ions},
  author={Ogorodnikova, OV and P{\l}oci{\'n}ski, T and Andrzejczuk, M and Rasi{\'n}ski, M and Mayer, M and Kurzyd{\l}owski, KJ and others},
  journal={Nuclear Instruments and Methods in Physics Research Section B: Beam Interactions with Materials and Atoms},
  volume={317},
  pages={159--164},
  year={2013},
  publisher={Elsevier}
}

@article{zenobia2012response,
  title={The response of polycrystalline tungsten to 30 keV helium ion implantation at normal incidence and high temperatures},
  author={Zenobia, Samuel J and Garrison, Lauren M and Kulcinski, Gerald L},
  journal={Journal of Nuclear Materials},
  volume={425},
  number={1-3},
  pages={83--92},
  year={2012},
  publisher={Elsevier}
}

@article{yi2018high,
  title={High-temperature damage evolution in 10 keV He+ irradiated W and W-5Re},
  author={Yi, Xiaoou and Arakawa, Kazuto and Ferroni, Francesco and Jenkins, Michael L and Han, Wentuo and Liu, Pingping and Wan, Farong},
  journal={Materials Characterization},
  volume={145},
  pages={77--86},
  year={2018},
  publisher={Elsevier}
}

@article{debelle2008first,
  title={First temperature stage evolution of irradiation-induced defects in tungsten studied by positron annihilation spectroscopy},
  author={Debelle, A and Barthe, MF and Sauvage, T},
  journal={Journal of Nuclear Materials},
  volume={376},
  number={2},
  pages={216--221},
  year={2008},
  publisher={Elsevier}
}

@article{mason2020observation,
  title={Observation of transient and asymptotic driven structural states of tungsten exposed to radiation},
  author={Mason, Daniel R and Das, Suchandrima and Derlet, Peter M and Dudarev, Sergei L and London, Andrew J and Yu, Hongbing and Phillips, Nicholas W and Yang, David and Mizohata, Kenichiro and Xu, Ruqing and others},
  journal={Physical Review Letters},
  volume={125},
  number={22},
  pages={225503},
  year={2020},
  publisher={APS}
}

@article{bonny2020trends,
  title={Trends in vacancy distribution and hardness of high temperature neutron irradiated single crystal tungsten},
  author={Bonny, G and Konstantinovic, MJ and Bakaeva, A and Yin, C and Castin, N and Mergia, K and Chatzikos, V and Dellis, S and Khvan, T and Bakaev, A and others},
  journal={Acta Materialia},
  volume={198},
  pages={1--9},
  year={2020},
  publisher={Elsevier}
}

@article{ogorodnikova2019annealing,
  title={Annealing of radiation-induced defects in tungsten: positron annihilation spectroscopy study},
  author={Ogorodnikova, OV and Dubov, L Yu and Stepanov, Sergey V and Terentyev, Dmitry and Funtikov, Yu V and Shtotsky, Yu V and Stolbunov, Valeriy S and Efimov, V and Gutorov, Konstantin},
  journal={Journal of Nuclear Materials},
  volume={517},
  pages={148--151},
  year={2019},
  publisher={Elsevier}
}

@article{chen2023towards,
  title={Towards a reliable nanohardness-dose correlation of ion-irradiated materials from nanoindentation tests: A case study in proton-irradiated vanadium},
  author={Chen, Shang and Yuan, Jiuxi and Wang, Shumin and Mei, Luyao and Yan, Jiaohui and Li, Lei and Zhang, Qiuhong and Zhu, Zhixi and Lv, Jin and Xue, Yunfei and others},
  journal={International Journal of Plasticity},
  volume={171},
  pages={103804},
  year={2023},
  publisher={Elsevier}
}

@article{abernethy2019effects,
  title={Effects of neutron irradiation on the brittle to ductile transition in single crystal tungsten},
  author={Abernethy, RG and Gibson, JS and Giannattasio, A and Murphy, JD and Wouters, O and Bradnam, S and Packer, LW and Gilbert, MR and Klimenkov, M and Rieth, M and others},
  journal={Journal of Nuclear Materials},
  volume={527},
  year={2019},
  publisher={Elsevier Science BV}
}

@article{xu2003molecular,
  title={Molecular dynamics simulation of vacancy diffusion in tungsten induced by irradiation},
  author={Xu, Q and Yoshiie, T and Huang, HC},
  journal={Nuclear Instruments and Methods in Physics Research Section B: Beam Interactions with Materials and Atoms},
  volume={206},
  pages={123--126},
  year={2003},
  publisher={Elsevier}
}

@article{saha2023microstructure,
  title={Microstructure and defect evolution in oxygen ion-irradiated pure nickel--Insights from experimental probes and molecular dynamics simulations},
  author={Saha, Uttiyoarnab and Dutta, Argha and Konkati, Chethan and Chakraborty, Soumita and Dey, Santu and Chauhan, Ankur and Srivastava, Sachin and Gayathri, N and Mukherjee, P},
  journal={Materials Chemistry and Physics},
  volume={305},
  pages={127916},
  year={2023},
  publisher={Elsevier}
}

@article{guo2009pressure,
  title={Pressure of stable He--vacancy complex in bcc iron: Molecular dynamics simulations},
  author={Guo, SH and Zhu, BE and Liu, WC and Pan, ZY and Wang, YX},
  journal={Nuclear Instruments and Methods in Physics Research Section B: Beam Interactions with Materials and Atoms},
  volume={267},
  number={18},
  pages={3278--3281},
  year={2009},
  publisher={Elsevier}
}

@article{rodney1999dislocation,
  title={Dislocation pinning by small interstitial loops: a molecular dynamics study},
  author={Rodney, D and Martin, G},
  journal={Physical review letters},
  volume={82},
  number={16},
  pages={3272},
  year={1999},
  publisher={APS}
}

@article{makin1968obstacles,
  title={The obstacles responsible for the hardening of neutron irradiated copper crystals},
  author={Makin, MJ},
  journal={Philosophical Magazine},
  volume={18},
  number={156},
  pages={1245--1255},
  year={1968},
  publisher={Taylor \& Francis}
}

@article{makin1965model,
  title={A model of “lattice” hardening in irradiated copper crystals with the external characteristics of “source” hardening},
  author={Makin, MJ and Sharp, JV},
  journal={physica status solidi (b)},
  volume={9},
  number={1},
  pages={109--118},
  year={1965},
  publisher={Wiley Online Library}
}

@article{garrison2019mechanical,
  title={Mechanical properties of single-crystal tungsten irradiated in a mixed spectrum fission reactor},
  author={Garrison, Lauren M and Katoh, Yutai and Kumar, NAP Kiran},
  journal={Journal of Nuclear Materials},
  volume={518},
  pages={208--225},
  year={2019},
  publisher={Elsevier}
}

@article{zheng2010investigation,
  title={Investigation of radiation damage in stainless steel, tungsten and tantalum by heavy ion irradiations},
  author={Zheng, Yongnan and Zuo, Yi and Yuan, Daqing and Zhou, Dongmei and Xu, Yongjun and Fan, Ping and Zhu, Jiazheng and Wang, Zhiqiang and Zhu, Shengyun},
  journal={Nuclear Physics A},
  volume={834},
  number={1-4},
  pages={761c--763c},
  year={2010},
  publisher={Elsevier}
}

@article{lee2007hydrogen,
  title={Hydrogen and helium trapping in tungsten under simultaneous irradiations},
  author={Lee, HT and Haasz, AA and Davis, JW and Macaulay-Newcombe, RG and Whyte, DG and Wright, GM},
  journal={Journal of nuclear materials},
  volume={363},
  pages={898--903},
  year={2007},
  publisher={Elsevier}
}

@article{lloyd2019decoration,
  title={Decoration of voids with rhenium and osmium transmutation products in neutron irradiated single crystal tungsten},
  author={Lloyd, Matthew J and Abernethy, Robert G and Gilbert, Mark R and Griffiths, Ian and Bagot, Paul AJ and Nguyen-Manh, Duc and Moody, Michael P and Armstrong, David EJ},
  journal={Scripta Materialia},
  volume={173},
  pages={96--100},
  year={2019},
  publisher={Elsevier}
}

@article{ren2022revealing,
  title={Revealing the synergistic effect of invisible helium clusters in helium irradiation hardening in tungsten},
  author={Ren, Qing-Yuan and Li, Yu-Hao and Gao, Ning and Han, Wei-Zhong and Niu, Yu-Ze and Xie, Hong-Xian and Zhang, Ying and Gao, Fei and Lu, Guang-Hong and Zhou, Hong-Bo},
  journal={Scripta Materialia},
  volume={219},
  pages={114850},
  year={2022},
  publisher={Elsevier}
}

@article{ren2025unveiling,
  title={Unveiling the inhibitory effect of hydrogen-decorated voids and dislocation loops on the glide of edge dislocation in tungsten},
  author={Ren, Qing-Yuan and Li, Yu-Hao and Du, Yu-Chen and Yang, Tian-Ren and Terentyev, Dmitry and Han, Wei-Zhong and Zhou, Hong-Bo and Lu, Guang-Hong},
  journal={Nuclear Fusion},
  volume={65},
  number={2},
  pages={026044},
  year={2025},
  publisher={IOP Publishing}
}

@article{yang2023influence,
  title={Influence of rhenium-decorated dislocation loops on edge dislocation gliding in tungsten},
  author={Yang, Tian-Ren and Li, Yu-Hao and Ren, Qing-Yuan and Terentyev, Dmitry and Xie, Hong-Xian and Gao, Ning and Zhou, Hong-Bo and Gao, Fei and Lu, Guang-Hong},
  journal={Scripta Materialia},
  volume={235},
  pages={115624},
  year={2023},
  publisher={Elsevier}
}

@article{osetsky2021atomic,
  title={Atomic-scale mechanisms of void strengthening in tungsten},
  author={Osetsky, Yuri N},
  journal={Tungsten},
  volume={3},
  number={1},
  pages={65--71},
  year={2021},
  publisher={Springer}
}

@article{birosca2019dislocation,
  title={The dislocation behaviour and GND development in a nickel based superalloy during creep},
  author={Birosca, Soran and Liu, Gang and Ding, Rengen and Jiang, Jun and Simm, Thomas and Deen, Chris and Whittaker, Mark},
  journal={International Journal of Plasticity},
  volume={118},
  pages={252--268},
  year={2019},
  publisher={Elsevier}
}

@article{hu2025new,
  title={New Insight into the Quantifying Vacancy Distribution in Self-Ion-Irradiated Tungsten: A Combined Experimental and Computational Study},
  author={Hu, Zhiwei and Wu, Jintong and Yang, Qigui and Jomard, Fran{\c{c}}ois and Granberg, Fredric and Barthe, Marie-France},
  journal={Nano Letters},
  volume={25},
  number={27},
  pages={10787--10793},
  year={2025},
  publisher={ACS Publications}
}

@article{edwards2022mapping,
  title={Mapping pure plastic strains against locally applied stress: Revealing toughening plasticity},
  author={Edwards, Thomas EJ and Maeder, Xavier and Ast, Johannes and Berger, Luisa and Michler, Johann},
  journal={Science advances},
  volume={8},
  number={30},
  pages={eabo5735},
  year={2022},
  publisher={American Association for the Advancement of Science}
}

@article{chatzikos2022positron,
  title={Positron annihilation spectroscopy investigation of defects in neutron irradiated tungsten materials},
  author={Chatzikos, Vasileios and Mergia, Konstantina and Bonny, Giovanni and Terentyev, Dmitry and Papadakis, Dimitrios and Pavlou, GE and Messoloras, Spyros},
  journal={International Journal of Refractory Metals and Hard Materials},
  volume={105},
  pages={105838},
  year={2022},
  publisher={Elsevier}
}

@article{xu2021recent,
  title={Recent progress of radiation response in nanostructured tungsten for nuclear application},
  author={Xu, Hang and He, Lan-Li and Pei, Yong-Feng and Jiang, Chang-Zhong and Li, Wen-Qing and Xiao, Xiang-Heng},
  journal={Tungsten},
  volume={3},
  number={1},
  pages={20--37},
  year={2021},
  publisher={Springer}
}

@article{li2021revealing,
  title={Revealing nano-scale lattice distortions in implanted material with 3D Bragg ptychography},
  author={Li, Peng and Phillips, Nicholas W and Leake, Steven and Allain, Marc and Hofmann, Felix and Chamard, Virginie},
  journal={Nature communications},
  volume={12},
  number={1},
  pages={7059},
  year={2021},
  publisher={Nature Publishing Group UK London}
}

@article{hasegawa2014neutron,
  title={Neutron irradiation effects on tungsten materials},
  author={Hasegawa, Akira and Fukuda, Makoto and Nogami, Shuhei and Yabuuchi, Kiyohiro},
  journal={Fusion Engineering and Design},
  volume={89},
  number={7-8},
  pages={1568--1572},
  year={2014},
  publisher={Elsevier}
}

@article{bonny2013mobility,
  title={On the mobility of vacancy clusters in reduced activation steels: an atomistic study in the Fe--Cr--W model alloy},
  author={Bonny, Giovanni and Castin, Nicolas and Bullens, J and Bakaev, Alexander and Klaver, TCP and Terentyev, Dimitri},
  journal={Journal of Physics: Condensed Matter},
  volume={25},
  number={31},
  pages={315401},
  year={2013},
  publisher={IOP Publishing}
}

@article{tanno2007effects,
  title={Effects of transmutation elements on neutron irradiation hardening of tungsten},
  author={Tanno, Takashi and Hasegawa, Akira and He, Jian-Chao and Fujiwara, Mitsuhiro and Nogami, Shuhei and Satou, Manabu and Shishido, Toetsu and Abe, Katsunori},
  journal={Materials Transactions},
  volume={48},
  number={9},
  pages={2399--2402},
  year={2007},
  publisher={The Japan Institute of Metals and Materials}
}

@article{yi2016insitu,
  title={In-situ TEM studies of 150 keV W+ ion irradiated W and W-alloys: Damage production and microstructural evolution},
  author={Yi, Xiaoou and Jenkins, Michael L and Kirk, Marquis A and Zhou, Zhongfu and Roberts, Steven G},
  journal={Acta Materialia},
  volume={112},
  pages={105--120},
  year={2016},
  publisher={Elsevier}
}

@article{armstrong2013effects,
  title={Effects of sequential tungsten and helium ion implantation on nano-indentation hardness of tungsten},
  author={Armstrong, DEJ and Edmondson, PD and Roberts, SG},
  journal={Applied Physics Letters},
  volume={102},
  number={25},
  year={2013},
  publisher={AIP Publishing}
}

@article{fikar2018nano,
  title={Nano-sized prismatic vacancy dislocation loops and vacancy clusters in tungsten},
  author={Fikar, Jan and Sch{\"a}ublin, Robin and Mason, Daniel R and Nguyen-Manh, Duc},
  journal={Nuclear Materials and Energy},
  volume={16},
  pages={60--65},
  year={2018},
  publisher={Elsevier}
}

@article{ghafarollahi2021theory,
  title={Theory of kink migration in dilute BCC alloys},
  author={Ghafarollahi, A and Curtin, WA},
  journal={Acta Materialia},
  volume={215},
  pages={117078},
  year={2021},
  publisher={Elsevier}
}

@article{osetsky2004dynamic,
  title={Dynamic properties of edge dislocations decorated by interstitial loops in $\alpha$-iron and copper},
  author={Osetsky, Yu N and Bacon, DJ and Rong, Z and Singh, BN},
  journal={Philosophical magazine letters},
  volume={84},
  number={11},
  pages={745--754},
  year={2004},
  publisher={Taylor \& Francis}
}

@article{lloyd2024microstructural,
  title={Microstructural evolution and transmutation in tungsten under ion and neutron irradiation},
  author={Lloyd, Matthew J and Haley, Jack and Jim, Bethany and Abernethy, Robert and Gilbert, Mark R and Martinez, Enrique and Hattar, Khalid and El-Atwani, Osman and Nguyen-Manh, Duc and Moody, Michael P and others},
  journal={Materialia},
  volume={33},
  pages={101991},
  year={2024},
  publisher={Elsevier}
}

@article{knaster2016materials,
  title={Materials research for fusion},
  author={Knaster, Juan and Moeslang, Anton and Muroga, Tatuo},
  journal={Nature Physics},
  volume={12},
  number={5},
  pages={424--434},
  year={2016},
  publisher={Nature Publishing Group UK London}
}

@article{das2018effect,
  title={The effect of helium implantation on the deformation behaviour of tungsten: X-ray micro-diffraction and nanoindentation},
  author={Das, S and Armstrong, DEJ and Zayachuk, Y and Liu, W and Xu, R and Hofmann, Felix},
  journal={Scripta Materialia},
  volume={146},
  pages={335--339},
  year={2018},
  publisher={Elsevier}
}

@article{das2018helium,
  title={Helium-implantation-induced lattice strains and defects in tungsten probed by X-ray micro-diffraction},
  author={Das, Suchandrima and Liu, W and Xu, R and Hofmann, Felix},
  journal={Materials \& Design},
  volume={160},
  pages={1226--1237},
  year={2018},
  publisher={Elsevier}
}

@article{fang2018hydrogen,
  title={Hydrogen embrittlement of tungsten induced by deuterium plasma: Insights from nanoindentation tests},
  author={Fang, Xufei and Kreter, Arkadi and Rasinski, Marcin and Kirchlechner, Christoph and Brinckmann, Steffen and Linsmeier, Christian and Dehm, Gerhard},
  journal={Journal of materials research},
  volume={33},
  number={20},
  pages={3530--3536},
  year={2018},
  publisher={Springer}
}

@article{reza2020thermal,
  title={Thermal diffusivity degradation and point defect density in self-ion implanted tungsten},
  author={Reza, Abdallah and Yu, Hongbing and Mizohata, Kenichiro and Hofmann, Felix},
  journal={Acta Materialia},
  volume={193},
  pages={270--279},
  year={2020},
  publisher={Elsevier}
}

@article{yi2013insitu,
  title={In situ study of self-ion irradiation damage in W and W--5Re at 500 C},
  author={Yi, X and Jenkins, ML and Briceno, M and Roberts, SG and Zhou, Z and Kirk, MA},
  journal={Philosophical Magazine},
  volume={93},
  number={14},
  pages={1715--1738},
  year={2013},
  publisher={Taylor \& Francis}
}

@article{wang2004orientation,
  title={Orientation dependence of nanoindentation pile-up patterns and of nanoindentation microtextures in copper single crystals},
  author={Wang, Yanwen and Raabe, Dierk and Kl{\"u}ber, Christian and Roters, Franz},
  journal={Acta materialia},
  volume={52},
  number={8},
  pages={2229--2238},
  year={2004},
  publisher={Elsevier}
}

@article{wang2020orientation,
  title={Orientation-dependent irradiation hardening in pure Zr studied by nanoindentation, electron microscopies, and crystal plasticity finite element modeling},
  author={Wang, Qiang and Cochrane, Christopher and Skippon, Travis and Wang, Zhouyao and Abdolvand, Hamidreza and Daymond, Mark R},
  journal={International Journal of Plasticity},
  volume={124},
  pages={133--154},
  year={2020},
  publisher={Elsevier}
}

@article{debroglie2015temperature,
  title={Temperature dependence of helium-implantation-induced lattice swelling in polycrystalline tungsten: X-ray micro-diffraction and Eigenstrain modelling},
  author={De Broglie, I and Beck, CE and Liu, W and Hofmann, Felix},
  journal={Scripta Materialia},
  volume={107},
  pages={96--99},
  year={2015},
  publisher={Elsevier}
}

@article{guan2017crystal,
  title={Crystal plasticity modelling and HR-DIC measurement of slip activation and strain localization in single and oligo-crystal Ni alloys under fatigue},
  author={Guan, Yongjun and Chen, Bo and Zou, Jinwen and Britton, T Ben and Jiang, Jun and Dunne, Fionn PE},
  journal={International Journal of Plasticity},
  volume={88},
  pages={70--88},
  year={2017},
  publisher={Elsevier}
}

@article{rieth2013recent,
  title={Recent progress in research on tungsten materials for nuclear fusion applications in Europe},
  author={Rieth, Michael and Dudarev, Sergei L and De Vicente, SM Gonzalez and Aktaa, J{\"u}rgen and Ahlgren, T and Antusch, S and Armstrong, DEJ and Balden, M and Baluc, Nadine and Barthe, M-F and others},
  journal={Journal of Nuclear Materials},
  volume={432},
  number={1-3},
  pages={482--500},
  year={2013},
  publisher={Elsevier}
}

@article{abernethy2017predicting,
  title={Predicting the performance of tungsten in a fusion environment: a literature review},
  author={Abernethy, RG},
  journal={Materials Science and Technology},
  volume={33},
  number={4},
  pages={388--399},
  year={2017},
  publisher={SAGE Publications Sage UK: London, England}
}

@article{wei2014first,
  title={First-principles study of the phase stability and the mechanical properties of W-Ta and W-Re alloys},
  author={Wei, Ning and Jia, Ting and Zhang, Xiaoli and Liu, Ting and Zeng, Z and Yang, XiaoYu},
  journal={AIP Advances},
  volume={4},
  number={5},
  year={2014},
  publisher={AIP Publishing}
}

@article{gilbert2011neutron,
  title={Neutron-induced transmutation effects in W and W-alloys in a fusion environment},
  author={Gilbert, MR and Sublet, J-Ch},
  journal={Nuclear Fusion},
  volume={51},
  number={4},
  pages={043005},
  year={2011}
}

@article{gilbert2012integrated,
  title={An integrated model for materials in a fusion power plant: transmutation, gas production, and helium embrittlement under neutron irradiation},
  author={Gilbert, MR and Dudarev, SL and Zheng, S and Packer, LW and Sublet, J-Ch},
  journal={Nuclear Fusion},
  volume={52},
  number={8},
  pages={083019},
  year={2012},
  publisher={IOP Publishing and International Atomic Energy Agency}
}

@article{hammond2017helium,
  title={Helium, hydrogen, and fuzz in plasma-facing materials},
  author={Hammond, KD},
  journal={Materials Research Express},
  volume={4},
  number={10},
  pages={104002},
  year={2017},
  publisher={IOP Publishing}
}

@article{beck2013effect,
  title={Effect of alloy composition \& helium ion-irradiation on the mechanical properties of tungsten, tungsten-tantalum \& tungsten-rhenium for fusion power applications},
  author={Beck, Christian E and Roberts, Steve G and Edmondson, Philip D and Armstrong, David EJ},
  journal={MRS Online Proceedings Library (OPL)},
  volume={1514},
  pages={99--104},
  year={2013},
  publisher={Cambridge University Press}
}

@article{ito2014molecular,
  title={Molecular dynamics simulation of a helium bubble bursting on tungsten surfaces},
  author={Ito, Atsushi M and Yoshimoto, Yoshihide and Saito, Seiki and Takayama, Arimichi and Nakamura, Hiroaki},
  journal={Physica Scripta},
  volume={159},
  number={1},
  pages={014062},
  year={2014},
  publisher={IOP Publishing}
}

@article{lhuillier2011trapping,
  title={Trapping and release of helium in tungsten},
  author={Lhuillier, Pierre-Emile and Belhabib, Taieb and Desgardin, Pierre and Courtois, Blandine and Sauvage, Thierry and Barthe, Marie-France and Thomann, Anne-Lise and Brault, Pascal and Tessier, Yves},
  journal={Journal of Nuclear Materials},
  volume={416},
  number={1-2},
  pages={13--17},
  year={2011},
  publisher={Elsevier}
}

@article{sandoval2015competing,
  title={Competing kinetics and He bubble morphology in W},
  author={Sandoval, Luis and Perez, Danny and Uberuaga, Blas P and Voter, Arthur F},
  journal={Physical review letters},
  volume={114},
  number={10},
  pages={105502},
  year={2015},
  publisher={APS}
}

@article{hofmann2015non,
  title={Non-contact measurement of thermal diffusivity in ion-implanted nuclear materials},
  author={Hofmann, Felix and Mason, Daniel R and Eliason, Jeffrey K and Maznev, Alexei A and Nelson, Keith A and Dudarev, Sergei L},
  journal={Scientific reports},
  volume={5},
  number={1},
  pages={16042},
  year={2015},
  publisher={Nature Publishing Group UK London}
}

@article{das2019orientation,
  title={Orientation-dependent indentation response of helium-implanted tungsten},
  author={Das, Suchandrima and Yu, Hongbing and Tarleton, Edmund and Hofmann, Felix},
  journal={Applied Physics Letters},
  volume={114},
  number={22},
  year={2019},
  publisher={AIP Publishing}
}

@article{gilbert2015energy,
  title={Energy spectra of primary knock-on atoms under neutron irradiation},
  author={Gilbert, Mark R and Marian, Jaime and Sublet, J-Ch},
  journal={Journal of nuclear materials},
  volume={467},
  pages={121--134},
  year={2015},
  publisher={Elsevier}
}

@article{terentyev2013radiation,
  title={Radiation-induced strengthening and absorption of dislocation loops in ferritic Fe--Cr alloys: the role of Cr segregation},
  author={Terentyev, D. and Bakaev, A.},
  journal={Journal of Physics: Condensed Matter},
  volume={25},
  number={26},
  pages={265702},
  year={2013},
  publisher={IOP Publishing}
}

@article{kazakov2024interaction,
  title={Interaction of edge dislocations with voids in tungsten},
  author={Kazakov, A. and Babicheva, R. I. and Zinovev, A. and Terentyev, D. and Zhou, K. and Korznikova, E. A. and Dmitriev, S. V.},
  journal={Tungsten},
  volume={6},
  number={3},
  pages={633--646},
  year={2024},
  publisher={Springer}
}

@article{gilbert2008structure,
  title={Structure and metastability of mesoscopic vacancy and interstitial loop defects in iron and tungsten},
  author={Gilbert, M. R. and Dudarev, S. L. and Derlet, P. M. and Pettifor, D. G.},
  journal={Journal of Physics: Condensed Matter},
  volume={20},
  number={34},
  pages={345214},
  year={2008}
}

@article{grammatikopoulos2019simulation,
  title={Simulation of the interaction between an edge dislocation and <111> interstitial dislocation loops in $\alpha$-iron},
  author={Grammatikopoulos, P. and Bacon, D. J. and Osetsky, Y. N.},
  journal={Radiation Effects and Defects in Solids},
  volume={174},
  number={3-4},
  pages={329--338},
  year={2019},
  publisher={Taylor \& Francis}
}

@article{bacon2006computer,
  title={Computer simulation of reactions between an edge dislocation and glissile self-interstitial clusters in iron},
  author={Bacon, D. J. and Osetsky, Y. N. and Rong, Z.},
  journal={Philosophical Magazine},
  volume={86},
  number={25-26},
  pages={3921--3936},
  year={2006},
  publisher={Taylor \& Francis}
}

@article{terentyev2007effect,
  title={The effect of temperature and strain rate on the interaction between an edge dislocation and an interstitial dislocation loop in $\alpha$-iron},
  author={Terentyev, D. and Malerba, L. and Bacon, D. J. and Osetsky, Y. N.},
  journal={Journal of Physics: Condensed Matter},
  volume={19},
  number={45},
  pages={456211},
  year={2007}
}

@article{liu2008molecular,
  title={Molecular dynamics simulations of the interactions between screw dislocations and self-interstitial clusters in body-centered cubic Fe},
  author={Liu, X.-Y. and Biner, S. B.},
  journal={Scripta Materialia},
  volume={59},
  number={1},
  pages={51--54},
  year={2008},
  publisher={Elsevier}
}

@article{terentyev2010reactions,
  title={Reactions between a 1/2< 111> screw dislocation and< 100> interstitial dislocation loops in alpha-iron modelled at atomic scale},
  author={Terentyev, D. and Bacon, D. J. and Osetsky, Y. N.},
  journal={Philosophical magazine},
  volume={90},
  number={7-8},
  pages={1019--1033},
  year={2010},
  publisher={Taylor \& Francis}
}

@article{terentyev2013cr,
  title={Cr segregation on dislocation loops enhances hardening in ferritic Fe--Cr alloys},
  author={Terentyev, D. and Bergner, F. and Osetsky, Y.},
  journal={Acta materialia},
  volume={61},
  number={5},
  pages={1444--1453},
  year={2013},
  publisher={Elsevier}
}

@article{das2019hardening,
  title={Hardening and strain localisation in helium-ion-implanted tungsten},
  author={Das, Suchandrima and Yu, Hongbing and Tarleton, Edmund and Hofmann, Felix},
  journal={Scientific reports},
  volume={9},
  number={1},
  pages={18354},
  year={2019},
  publisher={Nature Publishing Group UK London}
}

@article{das2020modified,
  title={Modified deformation behaviour of self-ion irradiated tungsten: A combined nano-indentation, HR-EBSD and crystal plasticity study},
  author={Das, S. and Yu, H. and Mizohata, K. and Tarleton, E. and Hofmann, F.},
  journal={International Journal of Plasticity},
  volume={135},
  pages={102817},
  year={2020},
  publisher={Elsevier}
}

@article{yadav2025characterisation,
  title={Characterisation of Irradiation Damage in Fe--3Cr and Fe--5Cr: A Study on the Effects of Chromium Content and Temperature},
  author={Yadav, C. B. and London, A. J. and Tadi{\'c}, T. and Xu, R. and Liu, W. and Fazinic, S. and Das, S.},
  journal={Materials Science and Engineering: A},
  pages={149464},
  year={2025},
  publisher={Elsevier}
}

@article{gilbert2014comparative,
  title={Comparative assessment of material performance in demo fusion reactors},
  author={Gilbert, M. R. and Zheng, S. and Kemp, R. and Packer, L. W. and Dudarev, S. L. and Sublet, J.-Ch.},
  journal={Fusion Science and Technology},
  volume={66},
  number={1},
  pages={9--17},
  year={2014},
  publisher={Taylor \& Francis}
}

@article{roy2025crystal,
  title={Crystal Plasticity Modeling of Hardening and Creep in Ferritic-Martensitic Alloys Under Thermal and Irradiation Environments},
  author={Roy, V. and Khan, I. A. and Patra, A.},
  journal={International Journal of Plasticity},
  pages={104513},
  year={2025},
  publisher={Elsevier}
}

@article{das2019recent,
  title={Recent advances in characterising irradiation damage in tungsten for fusion power},
  author={Das, S.},
  journal={SN Applied Sciences},
  volume={1},
  number={12},
  pages={1614},
  year={2019},
  publisher={Springer}
}

@article{hardie2013effects,
  title={Effects of irradiation temperature and dose rate on the mechanical properties of self-ion implanted Fe and Fe--Cr alloys},
  author={Hardie, C. D. and Williams, C. A. and Xu, S. and Roberts, S. G.},
  journal={Journal of Nuclear Materials},
  volume={439},
  number={1-3},
  pages={33--40},
  year={2013},
  publisher={Elsevier}
}

@article{durrschnabel2021new,
  title={New insights into microstructure of neutron-irradiated tungsten},
  author={D{\"u}rrschnabel, M. and Klimenkov, M. and J{\"a}ntsch, U. and Rieth, M. and Schneider, H. C. and Terentyev, D.},
  journal={Scientific reports},
  volume={11},
  number={1},
  pages={7572},
  year={2021},
  publisher={Nature Publishing Group UK London}
}

@article{you2017clustering,
  title={Clustering of transmutation elements tantalum, rhenium and osmium in tungsten in a fusion environment},
  author={You, Yu-Wei and Kong, Xiang-Shan and Wu, Xuebang and Liu, CS and Fang, QF and Chen, JL and Luo, G-N},
  journal={Nuclear Fusion},
  volume={57},
  number={8},
  pages={086006},
  year={2017},
  publisher={IOP Publishing}
}

@article{hirel2015atomsk,
  title={Atomsk: A tool for manipulating and converting atomic data files},
  author={Hirel, P.},
  journal={Computer Physics Communications},
  volume={197},
  pages={212--219},
  year={2015},
  publisher={Elsevier}
}

@article{stukowski2010ovito,
  title={Visualization and analysis of atomistic simulation data with OVITO--the Open Visualization Tool},
  author={Stukowski, A.},
  journal={Modelling and simulation in materials science and engineering},
  volume={18},
  number={1},
  pages={015012},
  year={2010}
}

@article{das2024md,
  title={Dislocation pinning in helium-implanted tungsten: A molecular dynamics study},
  author={Das, S. and Sand, A. and Hofmann, F.},
  journal={Journal of Nuclear Materials},
  volume={601},
  pages={155293},
  year={2024},
  publisher={Elsevier}
}

@article{terentyev2008simulation,
  title={Simulation of the interaction between an edge dislocation and a< 1 0 0> interstitial dislocation loop in $\alpha$-iron},
  author={Terentyev, D. and Grammatikopoulos, P. and Bacon, D. J. and Osetsky, Y. N.},
  journal={Acta Materialia},
  volume={56},
  number={18},
  pages={5034--5046},
  year={2008},
  publisher={Elsevier}
}

@article{yu2024atomistic,
  title={Atomistic investigation of the interaction between an edge dislocation and 1/2< 111> interstitial dislocation loops in irradiated tungsten},
  author={Yu, P. and Liu, G. and Shen, Y.},
  journal={International Journal of Plasticity},
  volume={179},
  pages={104030},
  year={2024},
  publisher={Elsevier}
}

@article{yi2015characterisation,
  title={Characterisation of radiation damage in W and W-based alloys from 2 MeV self-ion near-bulk implantations},
  author={Yi, X. and Jenkins, M. L. and Hattar, K. and Edmondson, P. D. and Roberts, S. G.},
  journal={Acta Materialia},
  volume={92},
  pages={163--177},
  year={2015},
  publisher={Elsevier}
}

@article{hasanzadeh2018three,
  title={Three-dimensional scanning transmission electron microscopy of dislocation loops in tungsten},
  author={Hasanzadeh, S. and Sch{\"a}ublin, R. and D{\'e}camps, B. and Rousson, V. and Autissier, E. and Barthe, M. F. and H{\'e}bert, C.},
  journal={Micron},
  volume={113},
  pages={24--33},
  year={2018},
  publisher={Elsevier}
}

\end{document}